\documentclass[twoside,twocolumn,9pt]{article}
\usepackage{extsizes}
\usepackage[super,sort&compress,comma]{natbib} 
\usepackage[version=3]{mhchem}
\usepackage[left=1.5cm, right=1.5cm, top=1.785cm, bottom=2.0cm]{geometry}
\usepackage{balance}
\usepackage{times,mathptmx}
\usepackage{sectsty}
\usepackage{graphicx}
\usepackage{subcaption}
\usepackage{lastpage}
\usepackage[format=plain,justification=justified,singlelinecheck=false,font={stretch=1.125,small,sf},labelfont=bf,labelsep=space]{caption}
\usepackage{float}
\usepackage{fancyhdr}
\usepackage{fnpos}
\usepackage[english]{babel}
\addto{\captionsenglish}{%
  
}
\usepackage{array}
\usepackage{droidsans}
\usepackage{charter}
\usepackage[T1]{fontenc}
\usepackage[usenames,dvipsnames]{xcolor}
\usepackage{setspace}
\usepackage[compact]{titlesec}
\usepackage{hyperref}

\usepackage{epstopdf}

\definecolor{cream}{RGB}{222,217,201}

\begin{document}

\pagestyle{fancy}
\thispagestyle{plain}
\fancypagestyle{plain}{

\fancyhead[C]{\includegraphics[width=18.5cm]{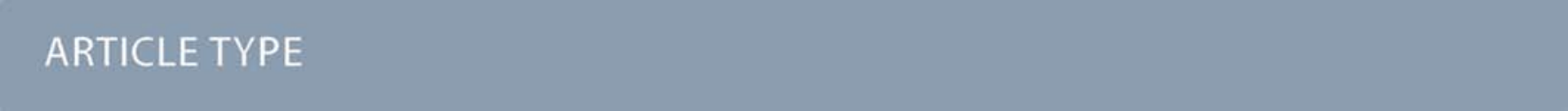}}
\fancyhead[L]{\hspace{0cm}\vspace{1.5cm}\includegraphics[height=30pt]{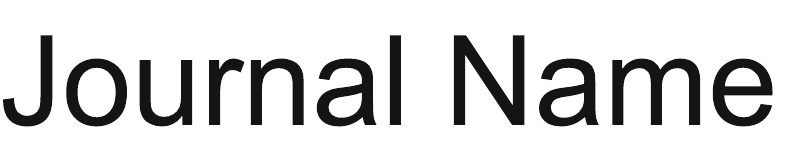}}
\fancyhead[R]{\hspace{0cm}\vspace{1.7cm}\includegraphics[height=55pt]{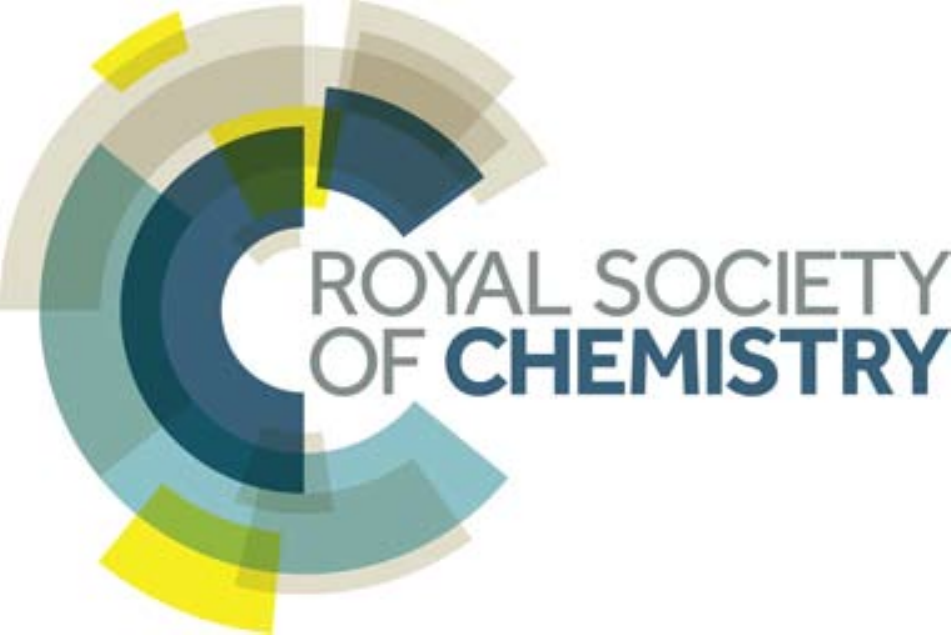}}
\renewcommand{\headrulewidth}{0pt}
}

\makeFNbottom
\makeatletter
\renewcommand\LARGE{\@setfontsize\LARGE{15pt}{17}}
\renewcommand\Large{\@setfontsize\Large{12pt}{14}}
\renewcommand\large{\@setfontsize\large{10pt}{12}}
\renewcommand\footnotesize{\@setfontsize\footnotesize{7pt}{10}}
\makeatother

\renewcommand{\thefootnote}{\fnsymbol{footnote}}
\renewcommand\footnoterule{\vspace*{1pt}%
\color{cream}\hrule width 3.5in height 0.4pt \color{black}\vspace*{5pt}} 
\setcounter{secnumdepth}{5}

\makeatletter 
\renewcommand\@biblabel[1]{#1}            
\renewcommand\@makefntext[1]%
{\noindent\makebox[0pt][r]{\@thefnmark\,}#1}
\makeatother 
\renewcommand{\figurename}{\small{Fig.}~}
\sectionfont{\sffamily\Large}
\subsectionfont{\normalsize}
\subsubsectionfont{\bf}
\setstretch{1.125} 
\setlength{\skip\footins}{0.8cm}
\setlength{\footnotesep}{0.25cm}
\setlength{\jot}{10pt}
\titlespacing*{\section}{0pt}{4pt}{4pt}
\titlespacing*{\subsection}{0pt}{15pt}{1pt}

\fancyfoot{}
\fancyfoot[LO,RE]{\vspace{-7.1pt}\includegraphics[height=9pt]{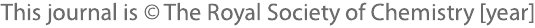}}
\fancyfoot[CO]{\vspace{-7.1pt}\hspace{13.2cm}\includegraphics{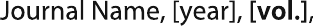}}
\fancyfoot[CE]{\vspace{-7.2pt}\hspace{-14.2cm}\includegraphics{head_foot/RF}}
\fancyfoot[RO]{\footnotesize{\sffamily{1--\pageref{LastPage} ~\textbar  \hspace{2pt}\thepage}}}
\fancyfoot[LE]{\footnotesize{\sffamily{\thepage~\textbar\hspace{3.45cm} 1--\pageref{LastPage}}}}
\fancyhead{}
\renewcommand{\headrulewidth}{0pt} 
\renewcommand{\footrulewidth}{0pt}
\setlength{\arrayrulewidth}{1pt}
\setlength{\columnsep}{6.5mm}
\setlength\bibsep{1pt}

\makeatletter 
\newlength{\figrulesep} 
\setlength{\figrulesep}{0.5\textfloatsep} 

\newcommand{\topfigrule}{\vspace*{-1pt}%
\noindent{\color{cream}\rule[-\figrulesep]{\columnwidth}{1.5pt}} }

\newcommand{\botfigrule}{\vspace*{-2pt}%
\noindent{\color{cream}\rule[\figrulesep]{\columnwidth}{1.5pt}} }

\newcommand{\dblfigrule}{\vspace*{-1pt}%
\noindent{\color{cream}\rule[-\figrulesep]{\textwidth}{1.5pt}} }

\makeatother

\twocolumn[
  \begin{@twocolumnfalse}
\vspace{3cm}
\sffamily
\begin{tabular}{m{4.5cm} p{13.5cm} }

\includegraphics{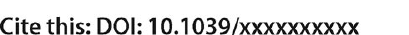} & \noindent\LARGE{\textbf{Amphoteric behavior of hydrogen ($H^{+1}$and $H^{-1}$) in complex hydrides from van der Waals interaction included \textit{ab-initio} calculation}} \\
\vspace{0.3cm} & \vspace{0.3cm} \\

 & \noindent\large{S Kiruthika,\textit{$^{a, b}$} H. Fjellv\aa g,\textit{$^{c}$} and P Ravindran $^{\ast}$\textit{$^{a,b}$}} \\

\includegraphics{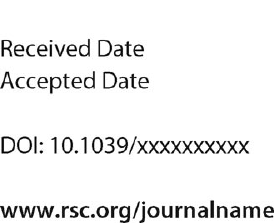} & \noindent\normalsize{In order to identify potential hydrogen storage materials, worldwide attention has been focused on hydrides with high gravimetric and volumetric capacity. Hydrogen is a unique element that possess positive, negative or neutral oxidation state in solids depending upon the chemical environment. If one can find hydrogen storage materials where hydrogen is present in both negative and positive oxidation state within the same structural framework then one can accommodate hydrogen with high volume density. So, it is fundamentally as well as technologically important to identify compounds in which hydrogen is in amphoteric nature and understand the necessary criteria for its origin. The experimental structural analysis of Cyclotriborazane and Diammonium dodeca hydro$-$closo$-$dodecaborate insinuate the presence of  hydrogen with anionic and cationic behavior within the same structure. In order to understand the role of van der Waals (vdW) interactions on structural parameters, we have considered 12 different vdW corrected functionals and found that the optPBE$-$vdW functional predicts the equilibrium structural parameters reliably with less than 0.009 \% accuracy. So, the optPBE-vdW functional is used to calculate charge density, electron localization function, total as well as the partial density of state, Bader and Born effective charge etc. We have observed that hydrogen is exhibiting amphoteric behavior with H closer to B in the negatively charged state and that neighboring to N is in a positively charged state.}\\ 
\end{tabular}

 \end{@twocolumnfalse} \vspace{0.6cm}
                                                                                                                                                                   ]

\renewcommand*\rmdefault{bch}\normalfont\upshape
\rmfamily
\section*{}
\vspace{-1cm}

\footnotetext{\textit{$^{a}$~Department of Physics, School of Basics and Applied Science, Central University of Tamil Nadu, Thiruvarur, India. Tel: 8300178007; E-mail:raviphy@cutn.ac.in}}
\footnotetext{\textit{$^{b}$~Simulation Center for Atomic and Nanoscale MATerials (SCANMAT), Central University of Tamil Nadu, Thiruvarur, India.}}
\footnotetext{\textit{$^{c}$~Center for Materials Science and Nanotechnology and Department of Chemistry, University of Oslo, Box 1033 Blindern, N0315, Norway.}}




\section{Introduction}
In the present scenario, hydrogen energy is considered as one of the environmental friendly renewable energy technologies which can provide electricity and fuel for various applications. Hydrogen is an ideal energy carrier, as it is non$-$polluting and gives up its electrons upon oxidation to form harmless water. Although it is the most abundant element in the universe, elemental hydrogen is not present in great quantities on earth. One of the most challenging problems in this field is to store hydrogen with a sufficient gravimetric as well as volumetric capacity. The overall requirements to use hydrogen storage materials for practical room$-$temperature applications are summarized as gravimetric capacity of \textgreater 6 wt\% and volumetric capacity \textgreater 50 Kg/m$^3$)~\cite{zuttel2007hydrogen,zuttel2003materials, jena2011materials, sakintuna2007metal} apart from good kinetics, reversibility, and desorption temperature to be just above room temperature. To overcome the technical limitation associated with storing hydrogen efficiently, recent research has focused on solid-state storage through chemical interactions, which is a viable approach. The solid-state storage materials have got tremendous attention since it is the only solution to achieve the required gravimetric and volumetric density. The solid$-$state storage materials include metal hydrides~\cite{schlapbach2011hydrogen}, intermetallic hydrides~\cite{matar2010intermetallic}, complex hydrides~\cite{ley2014complex}, and chemical hydrides~\cite{yu2017recent}. In comparison, complex hydrides have good hydrogen storage capacity with promising properties such as lightweight, high volume density, etc. But they lack reversibility and posses high decomposition temperature as well as slow kinetics those remain the problem to use them on a wide scale. However, Bogdanovi{\'c} and Schwickardi~\cite{bogdanovic1997ti} reported that the kinetic and reversible characteristics of complex hydrides can be increased by doping selected catalysts. \cite{schuth2004light, li2011recent, chen2002interaction}. Based on the Bogdanovi{\'c}, and Schwickardi~\cite{bogdanovic1997ti} suggestion Khan \textit{et al.} \cite{khan2016chloride} investigated the effect of doping of TiCl$_{3}$, Nb$_{2}$O$_{5}$, PbCl$_{2}$ and CeCl$_{3}$ on sodium alanate (NaAlH$_{4}$) towards its hydrogen storage properties. They concluded that doping various catalysts in NaAlH$_{4}$ enhances the reversibility and kinetics. Though the kinetics and reversible characteristics are improved by doping of selective catalysts in alanates, Resan \textit{et al.} \cite{resan2005effects} found that the following catalysts (TiCl$_4$ and TiCl$_3$) doping on lithium alanate (LiAlH$_{4}$) results in decreased amount of desorbed hydrogen due to the non$-$occurrence of the first step of hydrogen release. So, despite alanates have good hydrogen storage capacity, further improvement in hydrogen storage properties are needed to use them for commercial applications. 

On the other hand, borohydrides ([BH$_4$]$^-$) are also considered to be the promising candidates for hydrogen storage applications. Kim \textit{et al.}~\cite{kim2008reversible} reviewed the hydrogen sorption properties of TiCl$_{3}$ $-$catalyzed Ca(BH$_{4}$)$_{2}$ which was prepared by using high energy ball milling and found that the rehydrogenation conditions are alleviated compared to those of catalyzed doped LiBH$_4$. The above work clearly indicates that Ca(BH$_{4}$)$_{2}$ has considerable potential as the reversible hydrogen storage material. Also, Kim  \textit{et al.} suggested that more work is required to achieve full reversibility at more moderate conditions by adopting more effective catalysts or solid$-$state reactions. Inevitably, Rude \textit{et al.} ~\cite{rude2011tailoring} reviewed about the ways to improve the kinetic and thermodynamic properties in borohydrides and recommended the following (i) anion substitution, (ii) reactive hydride composites, and (iii) nano$-$confinement of hydrides and chemical reactions. Finally, Rude \textit{et al.} proposed that the anion substitution may lead to new solid solutions. Reactions between different hydrides are an efficient way to tailor the reaction thermodynamics. Nano$-$confinement is an emerging powerful technique for tailoring kinetic and thermodynamic properties of chemical reactions for a variety of purposes. In order to design novel borohydrides, we have to select a suitable metals that may tailor the hydrogen release temperature. From the above observations, it is clear that the overall requirements for efficient hydrogen storage are not yet formulated.

Amides and Imides are also classified under the complex hydrides. Lithium amides/imide materials have the presumption over the storage of hydrogen because of their ability to promote condensation reactions such as the introduction of amino groups into a molecule and removal of constituents of water.\cite{dafert1910new, ruff1911lithium} Chen \textit{et al.} \cite{chen2002interaction} first declared that Li$-$N$-$H and Ca$-$N$-$H systems provide feasibility for reversible hydrogen storage. However, to meet practical applications at more moderate temperatures with improved chemical stability, further works are needed regarding better material design, engineering, and mechanistic understanding. Later Luo \cite{luo2004linh2} proclaimed that (LiNH$_2$ $-$ MgH$_2$) is a viable hydrogen storage system after partial substitution of Li by Mg and it destabilizes lithium amide hydride system significantly according to the thermodynamic investigation on this system. LiNH$_2$ $-$ MgH$_2$ is a new hydrogen storage material, which can absorb hydrogen reversibly at a hydrogen pressure of 32 bar and temperature 200$^{\circ}$ C. This material is promising for on$-$board hydrogen storage, especially for fuel cells in vehicle transport. Hui wu\cite{wu2008structure} studied the hydrogen desorption and absorption in amide$-$metal hydride mixture and found that the dehydrogenation mechanism involves mobile small ions in both amide and metal hydrides and also explained the hydrogen storage mechanism for the ternary imide Li$_2$Ca(NH)$_2$. These results demonstrate that the mobility of small ions in the mixed amide/metal hydride system has a great impact on the hydrogen storage properties. 

Simultaneously, ammonia has the greater attraction towards the energy storage applications due to its ability to store high weight percentage of hydrogen. Klerke \textit{et al.}~\cite{klerke2008ammonia} briefly discussed about ammonia as a hydrogen carrier. The presence of ammonia in metal amine salts maintains the high volumetric hydrogen density and in addition to this, it is an attractive alternative which can solve many issues for hydrogen storage materials such as fast kinetics, high hydrogen storage capability, high availability, and low cost. Its less attractive side as a hydrogen carrier include the current methane$-$based production of ammonia without carbon sequestration and the toxicity of the liquid. 

Corfield \textit{et al.}~\cite{corfield1973crystal} first discussed about the crystal structure of cyclotriborazane (BH$_2$NH$_2$)$_3$ in 1972. Similarly, Yang \textit{et al.}~\cite{yang2008structural} studied the crystal structure of fully deuterated BH$_3$NH$_3$ using neutron diffraction and revealed the differences in the crystal structure of BD$_3$ND$_3$ from BH$_3$NH$_3$.~\cite{dalebrook2013hydrogen,orimo2007complex,van2007electronic,
li2011research,george2010structural,y2004first} Anion complexes present in complex hydrides prefers a negative oxidation state for hydrogen. However, in most of the other hydrogen storage materials hydrogen is present in positive oxidation state (H$^{+1}$) and based on structural analysis we have found that more than 85\% of the known hydrides contain hydrogen in positive oxidation state. ~\cite{peng2008ammonia,bluhm2006amineborane,miranda2007ab,li2014ammonia,lin2014high}
Further, if all the hydrogen in a solid exhibit same oxidation state, due to the Coulomb repulsion between hydrogen in same charge state, they stay apart from each other and hence the interatomic distance between the hydrogen ion in metals is 2\,  \AA\,or more with few exceptions.~\cite{ravindran2002violation,vajeeston2004search} This limits the volumetric density of hydrogen in solids. 

If one identifies the hydrides in which hydrogen is in both negative and positive oxidation states, then due to Coulomb attraction of H$^{+}$ and H$^{-}$ ions in such system, one can increase their gravimetric as well as the volumetric hydrogen density. Moreover, if one understand the origin of the presence of amphoteric hydrogen in the same structural frame$-$work in stable condition, this will pave the way to design potential hydrogen storage materials. Typically, hydrogen  has the nature to form compound either with the electronegative elements (F, N, O, Br, and Cl) or with the electropositive elements (Li, Na, K, Mg etc.). Hence, the hydrogen is in the negative or positive oxidation state alone within the same structural frame$-$work except in a few cases where it is believed to exist in both oxidation states. Based on experimental structural analysis it is concluded \cite{corfield1973crystal} that hydrogen is in both $+$1 and $-$1 oxidation states within the same compound sometime and two of such systems are considered in the present study. If the hydrogen is present in the amphoteric state within the same structural frame$-$work, it is of great fundamental interest and technological significance and hence, we have envisaged to carry out investigations on the chemical bonding between constituents in (BH$_2$NH$_2$)$_3$ and (NH$_4$)$_2$(B$_{12}$H$_{12}$) systems where it is interpreted that hydrogen is exists in amphoteric state. The theoretical understanding of the behavior of amphoteric hydrogen ($+$1 and $-$1) and its origin will helpful to design potential hydrogen storage materials with high volumetric density. For example, the volume density for (BH$_{2}$NH$_{2}$)$_{3}$ and (NH$_{4}$)$_{2}$(B$_{12}$H$_{12}$) are 134.50 g H$_{2}$/L and 130.18 g H$_{2}$/L, respectively and this value are higher than that of the state$-$of$-$the$-$art hydrogen storage materials such as MgH$_{2}$(109 g H$_{2}$/L), \cite{siegel2007thermodynamic} LiBH$_{4}$ (121 g H$_{2}$/L),\cite{zuttel2003hydrogen} NaAlH$_{4}$ (52 g H$_{2}$/L), \cite{yang2010high} Mg(NH$_{2}$)$_{2}$ (66 g H$_{2}$/L), \cite{xiong2005thermodynamic} and  NH$_{3} $BH$_{3}$ (96 g H$_{2}$/L) \cite{wolf2000calorimetric}. Due to the presence of high volumetric density in (BH$_{2}$NH$_{2}$)$_{3}$ and (NH$_{4}$)$_{2}$(B$_{12}$H$_{12}$), they can be considered as potential hydrogen storage materials. This motivated us to undertake the structural, electronic structure, chemical bonding, and total energy calculations for these compounds using a computational method based on density functional theory (DFT). 
\section{Computational Details}
We have used DFT~\cite{hohenberg1964inhomogeneous} within the generalized gradient approximation (GGA)~\cite{perdew1996generalized,perdew1986density} as implemented in the PAW method with plane wave basis set in the Vienna \textit{ab initio} simulations package(VASP)~\cite{kresse1993ab}. The structural optimization was continued until all atoms obtained their equilibrium positions as specified by Hellmann$-$Feynman forces of less than 10$^{-3}$ eV~\AA$^{-1}$ acting on the atoms. The Brillouin Zone integrations were performed with Monkhorst pack method for structural optimization and tetrahedron method to obtain density of states. All the calculations were performed with the $5\times5\times2$ and $4\times 4\times 4$ \textbf{k}$-$point mesh for (BH$_{2}$NH$_{2}$)$_{3}$ and (NH$_{4}$)$_{2}$(B$_{12}$H$_{12}$), respectively. In order to ensure well converged basis set, the plane wave cut$-$off of 400\,eV was used for both (BH$_{2}$ NH$_{2}$)$_{3}$ and (NH$_{4}$)$_{2}$(B$_{12}$H$_{12}$). The bonding, non$-$bonding, and anti$-$bonding states orginating from the bonding interaction between constituents were obtained using the Crystal Orbital Hamiltonian Population (COHP) analysis with LOBSTER package.~\cite{dronskowski1993crystal} Moreover, the Born effective charges and Bader effective charges were calculated utilizing the VASP package. To account the weak van der Waals interaction between molecule$-$like structural sub$-$units we have used vdW$-$corrected DFT method. In order to check the predicting capability of various vdW functionals, equilibrium bond lengths and dispersion behaviour of total energy curve were obtained using various van der Waals functionals implemented in the VASP code such as, revPBE$-$vdW, rPW86$-$vdW, optPBE $-$vdW, optB88$-$vdW, optB86b$-$vdW,~\cite{dion2004van,roman2009efficient,klimevs2009chemical,lee2010higher,klimevs2011van,thonhauser2007van} DFT$-$D2$-$vdW, DFT$-$D3$-$vdW, ~\cite{grimme2006semiempirical,grimme2010consistent,grimme2011effect} TS$-$vdW, TS+SCS$-$vdW, TS$-$HI$-$vdW,~\cite{tkatchenko2009accurate,tkatchenko2012accurate,bucko2013improved, buvcko2014extending, buvcko2016many} MBD$-$vdW and dDsC$-$vdW~\cite{ambrosetti2014long,steinmann2011comprehensive,steinmann2011generalized,becke2005exchange}. The total energy was calculated as a function of volume using force as well as stress minimization in each point to estimate the equilibrium volume and also check the dispersion behavior of total energy vs volume curves.
\paragraph{Theoretical models to account van der Waals interactions} 
There are several van der Waals functionals were introduced in density functional theory to account for van der Waals interactions in solids. Among the various functionals introduced, the van der Waals functional proposed by Dion \textit{et al.} ~\cite{dion2004van} for the exchange correlation term is 
\begin{equation}
E_{xc} = E^{GGA}_{x} + E_{c}^{LDA}+ E_{c}^{nl}
\end{equation}
where the exchange energy, E$_{x}^{GGA}$ is obtained from the GGA functional proposed by Perdew$-$Burke$-$Ernzerhof (PBE)~\cite{perdew1996generalized} and the correlation energy, E$_{c}^{LDA}$ is obtained from the local density approximation. The term E$_{c}^{nl}$ is the non$-$local energy term which accounts approximately for the non$-$local electron correlation effects. Rom{\'a}n$-$P{\'e}rez and Soler ~\cite{roman2009efficient} have replaced the double spatial integral into fast Fourier transforms, this speeds up the computational time.
In the D2 method of Grimme ~\cite{grimme2006semiempirical}, the correction term takes the form: 
\begin{equation}
\label{eq-2}
E_{disp} =-\frac{1}{2}\sum_{i=1}^{N_{at}}\sum_{j=1}^{N_{at}}\sum_{L}\prime \dfrac{C_{6ij}}{r^{6}_{ij,L}} f_{d,6} (r_{ij,L})
\end{equation} 
where the summations run over all atoms N$_{at}$ and all translations of the unit cell L = $(l_1, l_2, l_3)$, the prime indicates that $i\neq j$ for L=0, C$_{6ij}$ denotes the dispersion coefficients for the atom pair $ij$, r$_{ij}$, L is the distance between atom $i$ located in the reference cell L=0 and atom $j$ in the cell L, and the term f(r$_{ij}$) is a damping function whose role is to scale the force  field to minimize the contributions from interactions within typical bond distances ~\cite{grimme2006semiempirical}. In practice, the terms in eq \ref{eq-2} corresponds to interactions over distances longer than a certain suitably chosen cut-off radius, which contributes only negligibly to the E$_{disp}$ and can be ignored. Parameters, C$_{6ij}$ and R$_{0ij}$ are computed using the following combination rules:
\begin{equation}
C_{6ij} = \sqrt{C_{6ii} C_{5jj}}
\end{equation}
\begin{equation}
R_{0ij} = R_{0i}+R_{0j}.
\end{equation}
The values of C$_{6ii}$ and R$_{0i}$ are tabulated for each element and are insensitive to the particular chemical situation (for instance, C$_6$ for carbon in methane takes  exactly the same value as that for C in benzene within this approximation). It may be noted that in the original method of Grimme, Fermi$-$type damping function is used as follows:
\begin{equation}
f_{d,6}(r_{ij}) = \dfrac{S_6}{1+e^{-d(r_{ij}/(S_R R_{0ij})-1)}}
\end{equation}
where the global scaling parameter S$_6$ has been optimized for different DFT functionals such as PBE, BLYP and B3LYP takes the values 0.75, 1.2 and 1.05, respectively and the parameter S$_R$ is usually fixed as 1.00.
In the D3 correction method of Grimme \textit{et al.}~\cite{grimme2010consistent,grimme2011effect} the following vdW-energy expression is used :
\begin{equation}
E_{disp} = -\dfrac{1}{2}\sum_{i=1}^{N_{at}} \sum_{j=1}^{N_{at}}\sum_L \prime \Big(f_{d,6} (r_{ij,L}) \frac{C_{6ij}}{r^6_{ij,L}} + f_{d,8} (r_{ij,L}) \frac{C_{8ij}}{r^8_{ij,L}}\Big)
\end{equation}
Unlike in the D2 method, the dispersion coefficients, $C_{6ij}$ are geometry-dependent as they are adjusted on the basis of local  geometry (coordination number) around atoms \textit{i} and \textit{j}~\cite{grimme2010consistent}. In the zero damping D3 method (D3(zero)), damping of the following form is used:
\begin{equation}
f_{d,n} (r_{ij}) = \dfrac{S_n}{1+6(r_{ij}/(S_{R,n} R_{0ij}))^{-\alpha_n}}
\end{equation}
where $R_{0ij}$ = $\sqrt{\dfrac{C_{8ij}}{C_{6ij}}}$, the parameters of $\alpha_6$, $\alpha_8$, $S_{R,8}$ are  fixed to have the values of 14, 16, and 1, respectively and $S_6$, $S_8$, and $S_{R,6}$ are adjustable parameters whose  values depend on the choice of exchange$-$correlation functional. The dispersion term alone varies for other van der Waals functionals such as the TS$-$vdW~\cite{tkatchenko2009accurate}, TS+SCS$-$vdW\cite{tkatchenko2012accurate}, TS$-$HI$-$vdW~\cite{bucko2013improved, buvcko2014extending, buvcko2016many}, MBD$-$vdW~\cite{ambrosetti2014long} and dDsC$-$vdW ~\cite{steinmann2011comprehensive,steinmann2011generalized,becke2005exchange}. \\
The expression for the dispersion energy within the method of Tkatchenko and Scheffler (TS$-$vdW) is formally identical to that of the DFT$-$D2$-$vdw method eq \ref{eq-2}. The important difference is that the dispersion coefficients and damping function  are charge density dependent. Still the TS$-$vdW fails to describe the structural parameters and energetics of ionic solids. This problem can be solved by replacing the convectional Hirshfeld partitioning used to compute properties of interacting atoms by the iterative scheme proposed by Bultinck\cite{bultinck2007critical}. In such case, the dispersion energy (TS+SCS$-$vdW) is computed using  the same equation as in the original TS$-$vdW method but with corrected parameters C$^{SCS}_{6ii}$, $\alpha^{SCS}_i$, and $R^{SCS}_{0i}$. The many$-$body dispersion energy method (MBD rsSCS) of Tkatchenko \textit{et al.}~\cite{tkatchenko2012accurate,ambrosetti2014long} is based on the random phase expression of correlation energy. In this method, ploarizability, dispersion coefficients, charge, and charge$-$overlap between atoms in molecules or in solids are computed on the basis of a simplified exchange$-$hole dipole moment formalism, pioneered by Becke and Johnson \cite{becke2005exchange}.\\
Using the above 12 van der Waals functionals we have calculated the total energy as a function of volume for (BH$_{2}$NH$_{2}$)$_{3}$ as well as (NH$_4$)$_2$(B$_{12}$H$_{12}$) and the results were compared to identify the suitable van der Waals functional to study similar materials. 
\section{Results and Discussion}
\subsection{Structural Description}
The crystal structure of Cyclotriborazane (BH$_2$NH$_2$)$_3$ is orthorhombic with space group of Pbcm (space group no 57)~\cite{corfield1973crystal} as shown in Fig. 1(a). In this structure the boron and nitrogen atoms are arranged alternative to one another like a benzene ring and also each boron as well as nitrogen atoms are bonded to two hydrogen atoms. Similarly, Fig. 1(b) shows the crystal structure of Diammonium dodeca hydro$-$closo$-$dodecaborate (NH$_4$)$_2$(B$_{12}$H$_{12}$) which stablizes in the face center cubic (FCC) structure with the space group of Fm$\overline{3}$ (space group no. 202).~\cite{tiritiris2003dodekahydro} In this crystal structure each nitrogen atom is tetrahedrally bonded with four hydrogen atoms and form molecular$-$like structural sub$-$units as shown in Fig. 1(b). The boron atoms are covalently bonded with five neighboring boron atoms and also with one isolated hydrogen atom making a dodecohedron resulting isolated cage$-$like molecular structural sub$-$units. The structural analysis of these compounds shows that, all the borons are bonded to hydrogen in negative oxidation state H($-$1). Similarly, the nitrogens are bonded to hydrogen with positive oxidation states H($+$1). From the computationally optimized crystal structure using optPBE$-$vdW functional the estimated bond distance between H($-$1)$-$B is 1.2~\AA\,whereas, that between H($+$1)$-$N is 1.0~\AA (see Table \ref{tbl:1}). The bond distance between the H($-$1)$-$H($+$1) in these compounds ranges from 1.9 to 2.2~\AA.~\cite{boddeker1966boron, leavers1969heat, sun2011first} 

\begin{figure}[h]
  \includegraphics[scale=0.23]{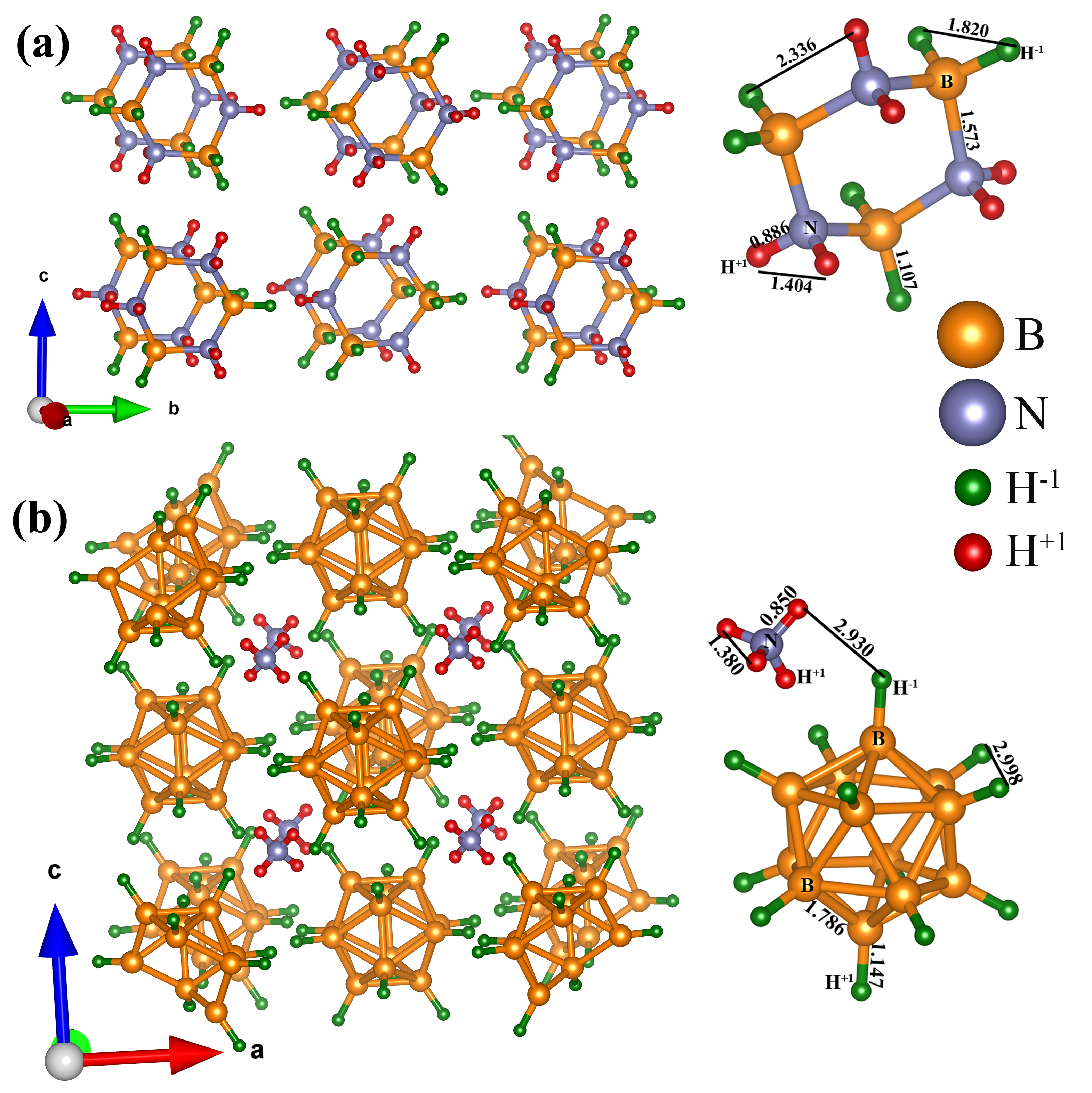}
  \caption{Crystal structure of (BH$_2$NH$_2$)$_3$ and (NH$_4$)$_2$(B$_{12}$H$_{12}$) are given in (a) and (b), respectively. The hydrogen with positive and negative oxidation states are depicted in red and green colour, respectively. The bond lengths are given in \AA}
  \label{fig:1}
\end{figure}

\begin{table}[h]
\small
\caption{The equilibrium bond length (in \AA) between constituents in (BH$_2$NH$_2$)$_3$ and (NH$_4$)$_2$(B$_{12}$H$_{12}$) obtained from Generalised Gradient Approximation(GGA) as well as van der Waals functional (optPBE$-$vdW) are compared with corresponding experimental values \cite{corfield1973crystal, tiritiris2003dodekahydro} }
\label{tbl:1}
\begin{tabular*}{0.48\textwidth}{@{\extracolsep{\fill}}ccccc}
\hline 
Compound Name & Bonds & \multicolumn{3}{c}{Bond Length(\AA)}\\
 & & Experimental & GGA & optPBE$-$vdW\\
\hline 
(BH$_2$ NH$_2$)$_3$ & H1$ - $B1 & 1.085 & 1.201 & 1.220\\
 & H2$ - $B1 & 1.107 & 1.221 & 1.218\\
 & H5$ - $B2 & 1.084 & 1.217 & 1.219\\
 & H6$ - $B2 & 1.116 & 1.218 & 1.220\\
 & H3$ - $N1 & 0.882 & 1.036 & 1.025\\
 & H4$ - $N1 & 0.886 & 1.029 & 1.024\\
 & H7$ - $N2 & 0.874 & 1.032 & 1.025\\
 & H8$ - $N2 & 0.886 & 1.035 & 1.025\\
\hline 
(NH$_4$)$_2$(B$_{12}$ H$_{12}$) & H1$-$B & 1.147 & 1.202 & 1.206\\
 & H2$ - $N & 0.848 & 1.035 & 1.040\\
\hline
\end{tabular*} 
\end{table}

\begin{table*}
\small
\caption{The optimized atom positions at the equilibrium volume obtained from generalized gradient approximation(GGA) and van der Waals functional (optPBE$-$vdW) for (BH$_2$NH$_2$)$_3$ and (NH$_4$)$_2$(B$_{12}$H$_{12}$)are compared with corresponding experimental values reported in \cite{corfield1973crystal, tiritiris2003dodekahydro}}
\label{tbl:2}
\begin{tabular*}{\textwidth}{@{\extracolsep{\fill}}lllll}
\hline
Compound (structure type; space group) & Wyckoff position & Experimental & GGA & optPBE$-$vdW \\ 
 & & & & \\ 
 & & & & \\
\hline
(BH$_2$ NH$_2$)$_3$;Pbcm & B1(4d)& 0.148, 0.115, 0.250 & 0.144, 0.119, 0.250 & 0.137, 0.117, 0.250\\
 & H1(4d) & 0.902, 0.118, 0.250 & 0.843, 0.114, 0.250 & 0.861, 0.122, 0.250 \\
 & H2(4d) & 0.248, 0.199, 0.250 & 0.236, 0.222, 0.250 & 0.251, 0.208, 0.250 \\                                     
 & B2(8e) & 0.150, 0.927, 0.368 & 0.154, 0.915, 0.376 & 0.139, 0.927, 0.369\\
 & H5(8e) & 0.904, 0.925, 0.372 & 0.854, 0.914, 0.373 & 0.863, 0.925, 0.373 \\
 & H6(8e) & 0.255, 0.886, 0.446 & 0.257, 0.863, 0.471 & 0.252, 0.881, 0.455 \\
 & N1(8e) & 0.254, 0.050, 0.362 & 0.267, 0.050, 0.373 & 0.243, 0.051, 0.363 \\
 & H3(8e) & 0.454, 0.053, 0.369 & 0.521, 0.058, 0.382 & 0.474, 0.056, 0.373 \\
 & H4(8e) & 0.184, 0.082, 0.427 & 0.175, 0.093, 0.453 & 0.159, 0.089, 0.438 \\
 & N2(4d) & 0.250, 0.871, 0.250 & 0.284, 0.855, 0.250 & 0.244, 0.869, 0.250 \\
 & H7(4d) & 0.454, 0.864, 0.250 & 0.539, 0.852, 0.250 & 0.474, 0.858, 0.250 \\
 & H8(4d) & 0.184, 0.803, 0.250 & 0.214, 0.765, 0.250 & 0.156, 0.791, 0.250 \\
\hline
(NH$_4$)$_2$(B$_{12}$ H$_{12}$);Fm$\overline{3}$  & B(48h) & 0.000, 0.132, 0.081 & 0.500, 0.134, 0.582 & 0.500, 0.132, 0.581\\
 & H1(48h) & 0.000, 0.223, 0.136 & 0.500, 0.230, 0.639 & 0.500, 0.227, 0.639 \\
 & N(8c)& 0.250, 0.250, 0.250 & 0.250, 0.250, 0.250 & 0.250, 0.250, 0.250 \\
 & H2(32f) & 0.205, 0.205, 0.205 & 0.194, 0.194, 0.194 & 0.194, 0.194, 0.194 \\
\hline
\end{tabular*}
\end{table*}

\begin{figure} [h]
\centering
 \includegraphics[scale=0.28]{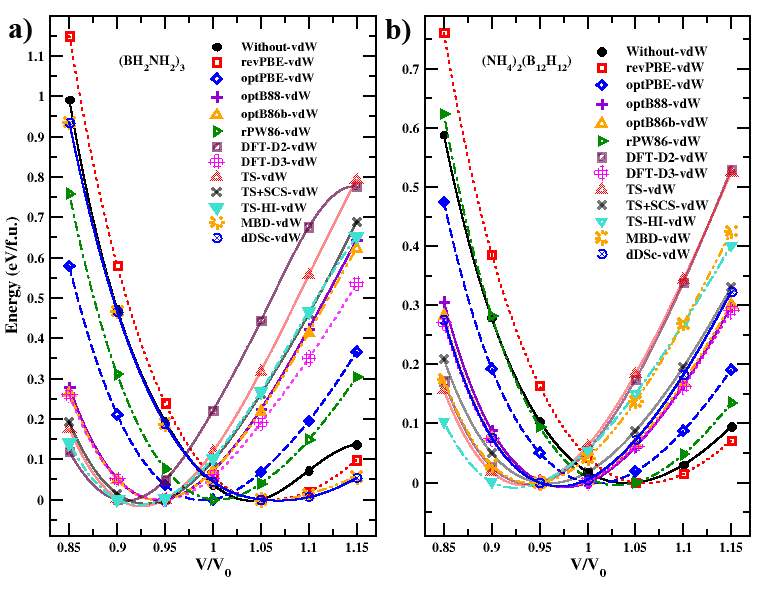}
 \caption{The total energy vs volume curve for (BH$_2$NH$_2$)$_3$ and (NH$_4$)$_2$(B$_{12}$H$_{12}$) are given in (a) and (b), respectively obtained from vdW$-$correction included functionals.}
 \label{fig:2}
\end{figure}

\begin{figure}[h]
\centering
 \includegraphics[scale=0.30]{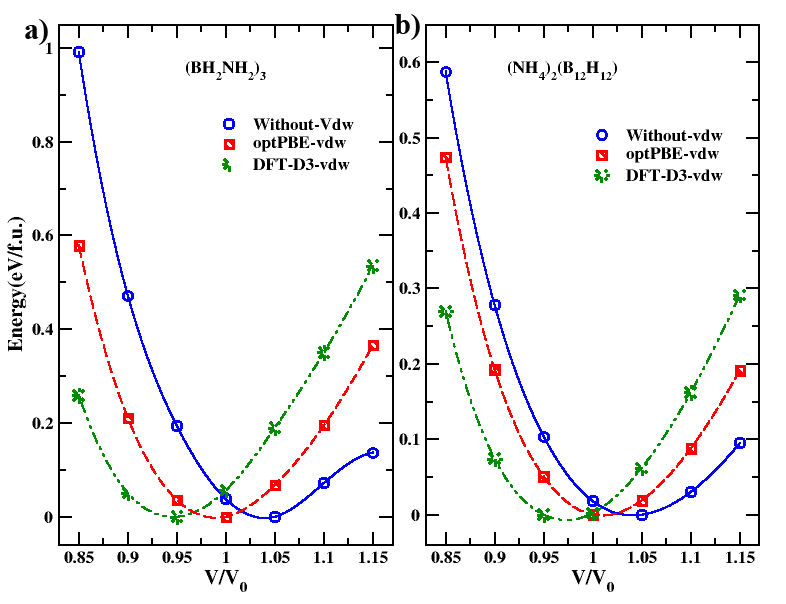}
 \caption{The total energy vs volume curves  for (BH$_2$NH$_2$)$_3$ and (NH$_4$)$_2$(B$_{12}$H$_{12}$) are given in (a) and (b), respectively obtained from \textit{ab}$-$inito calculation using GGA, optPBE$-$vdW, and DFT$-$D3 vdW$-$correction functionals.}
 \label{fig:3}
\end{figure}

\begin{table*}[h]
\small
\caption{Optimized equilibrium unit cell parameters (in \AA), equilibrium volume (in \AA $^{3}$)  and the deviation from experimental volume ($\delta$ V in \%) for (BH$_2$NH$_2$)$_3$ and (NH$_4$)$_2$(B$_{12}$H$_{12}$) obtained from GGA and various van der Waals interaction included functionals. The lattice parameters mentioned in brackets are from the coresponding experimental measurements \cite{corfield1973crystal,tiritiris2003dodekahydro}}
\label{tbl:3}
\begin{tabular*}{\textwidth}{@{\extracolsep{\fill}}ccccccccc}
\hline
\multicolumn{5}{c}{(BH$_2$NH$_2$)$_3$} & \multicolumn{4}{c}{(NH$_4$)$_2$(B$_{12}$H$_{12}$)} \\ 
vdW$-$DFT & Lattice parameter  & Cell volume  & $\delta$ V & Bulk modulus & Lattice parameter  & Cell volume  & $\delta$ V  & Bulk modulus \\ [1.5ex]
 & (\AA) & ({\AA}$^{3}$) & (\%) & (GPa) & (\AA) & ({\AA}$^{3}$) & (\%) & (GPa)\\
\hline
GGA & a=4.05(4.40)\cite{corfield1973crystal} & 630.04 & 4.38 & 8.40 & a=10.74(10.87)\cite{tiritiris2003dodekahydro} & 334.88 & 4.40 & 10.42 \\
 & b=11.08(12.21) & (603.57) & & & & (321.81) & &\\
 & c=10.32(11.22) & & & & & & &\\ 
revPBE$-$vdW & a=4.50 & 639.66 & 5.96 & 8.62 & a=7.84 & 341.01 & 5.6 & 11.10 \\
 & b=12.44 & & & & & & & \\
 & c=11.40 & & & & & & & \\ 
optPBE$-$vdW & a=4.40 & 597.28 & $-$1.05 & 10.82 & a=7.71 & 324.41 & 0.8 & 12.97 \\
 & b=12.16 & & & & & & & \\
 & c=11.14 & & & & & & & \\
optB88$-$vdW & a=4.33 & 568.20 & $-$5.87 & 12.74 & a=7.62 & 313.66 & $-$2.54 & 14.32 \\
 & b=11.96 & & & & &  & & \\
 & c=10.96 & & & & & & & \\
optB86b$-$vdW & a=4.32 & 567.71 & $-$5.95 & 12.32 & a=7.61 & 312.59 & $-$2.87 & 14.04 \\
 & b=11.96 &  &  & &  & & & \\
 & c=10.96 &  &  & &  & & & \\
rPW86$-$vdW & a=4.42 & 607.40 & 0.63 & 10.94 & a=7.76 & 331.49 &  3.00 & 12.80 \\
 & b=12.23 & & & &  & & &  \\
 & c=11.21 & & & &  & & & \\ 
DFT$-$D2$-$vdW & a = 4.33 & 545.78 & $-$9.58 & 16.76 & a=7.53 & 302.21 & $-$6.1 & 16.93  \\
 & b=11.97 & & & & & & & \\
 & c=10.97 & & & & & & & \\
DFT$-$D3$-$vdW & a=4.33 & 569.38 & $-$5.7 & 11.09 & a=7.61 & 312.73 & $-$2.83 & 13.43 \\
 & b=11.97 & & & & & & & \\
 & c=10.97 & & & & &  & & \\
TS$-$vdW & a=4.29 & 556.59 & $-$7.12 & 14.70 & a=7.51 & 300.60 &  $-$6.66 & 17.18 \\
 & b=11.92 & & & & & & & \\
 & c=10.92& & & & & & & \\
TS+SCS$-$vdW & a=4.29 & 560.60 & $-$7.12 & 12.61 & a=7.58 & 308.80 & $-$4.05 & 13.02 \\
 & b=12.08 &  &  &  &  & & &\\
 & c=11.02 &  &  &  &  & & & \\
TS$-$HI$-$vdW & a=4.27 & 555.35 &  $-$7.99 & 12.38 & a=7.50 & 298.39 & $-$7.28 & 13.70 \\
 & b=11.88 &  &  & & & & &  \\
 & c=10.93 &  &  & & &  & & \\ 
MBD$-$vdW & a=4.51 & 642.25 & 6.4 & 5.94& a=7.54 & 303.72 &  $-$5.63 & 14.93 \\
 & b=12.46 & & & & & & &  \\
 & c=11.42 & & & & & & & \\
dDSc$-$vdW & a=4.51 & 642.92 & 6.5 & 5.92 & a=7.61 & 312.08 & $-$3.03 & 14.33  \\
 & b=12.46 & & & & & & & \\
 & c=11.42 & & & & & & & \\
 \hline
\end{tabular*}
\end{table*}
We have optimized the atomic coordinates and unit cell parameters globally using stress as well as force minimization with various exchange correlation functionals, and thus identified the equilibrium cell volume, atomic coordinates, and unit cell dimensions which are listed in Table \ref{tbl:2} along with available experimental values. The optimized equilibrium structural parameters are obtained by varying the unit cell volume between $-$15\% to $+$15\% from experimental equilibrium volume with the step of 5\% and relaxed all atomic coordinates and unit cell dimensions globally for each volume. Fig. \ref{fig:2} illustrates the resulting total energy vs volume relation for the considered systems (BH$_2$NH$_2$)$_3$ and (NH$_4$)$_2$(B$_{12}$H$_{12}$) in Fig.~2(a) and 2(b), respectively. The equilibrium volume and bulk modulus were extracted from the calculated energy vs cell volume curves by fitting them to the universal equation of state proposed by Vinet \textit{et al.}~\cite{vinet1986universal,vinet1989universal} and are virtually the same when obtained by fitting to Brich~\cite{birch1947finite} and Murnaghan~\cite{murnaghan1944compressibility} equation of states. From the Fig. \ref{fig:2} we can observed that the total energy vs volume curve obtained from GGA calculation is (i.e. without vdW correction) overestimated with respect to experimental value by about 4.40\%. However, if one consider the van der Waals interaction to the calculation, the equilibrium volume obtained from total energy vs volume curve is deviated compared with experimental values by $-$9.58\% to 6.5 \% (see Table \ref{tbl:3}) depending upon the vdW functional we have used.\\  
The calculated total energy vs cell volume curve for $(BH_2NH_2)_3$ and (NH$_4$)$_2$(B$_{12}$H$_{12}$) obtained using the 12 vdW$-$functionals also given in Fig. \ref{fig:2}. Among this 12 functionals revPBE$-$vdW, rPW86$-$vdW, optPBE$-$vdW, optB88$-$vdW, and optB86b$-$vdW vdW are suggested based on vdW$-$DF2 of Langreth and Lundqvist with five different exchange correlation functionals. Among these five functionals, the rPW86$-$vdW functional predicted the equilibrium volume much closer to experiment (deviation 0.63\% for$(BH_2NH_2)_3$ and 3\% for (NH$_4$)$_2$(B$_{12}$H$_{12}$)) than that from all the other functionals. Followed by this, the optPBE$-$vdW functional give the equilibrium volume closer to experiment (deviating $-$1.05\% for (BH$_2$NH$_2$)$_3$ and 0.8\% for (NH$_4$)$_2$(B$_{12}$H$_{12}$)). On the other hand, the other seven vdW$-$ correction methods considered in the present study such as DFT$-$D2$-$vdW, DFT$-$D3$-$vdW, TS$-$vdW, TS+SCS$-$vdW, TS$-$HI$-$vdW, MBD$-$vdW, and dDSc$-$vdW; the DFT$-$D3$-$vdW shows lowest deviation in predicting equilibrium volume (deviating $-$5.7\% for (BH$_2$NH$_2$)$_3$, and $-$2.83 \% for (NH$_4$)$_2$(B$_{12}$H$_{12}$). So, for clarity we have displayed the total energy vs volume curves using optPBE$-$vdW and DFT$-$D3$-$vdW functionals along with that obtained  from the GGA calculation in Fig.~\ref{fig:3}. From these studies we found that rPW86$-$vdW and optPBE$-$vdW are more appropriate to predict structural properties of molecular$-$like hydrides such as (BH$_2$NH$_2$)$_3$ and (NH$_4$)$_2$(B$_{12}$H$_{12}$). However, optPBE$-$vdW yield overall good agreement with experimental equilibrium volume than other functionals for both systems and hence this was used for further analysis. Also, the 4.40\% overestimation of equilibrium volume by GGA calculation clearly indicates the importance of including van der Waals correction to predict the structural properties of these systems reliably.
\subsection{Density of States Analysis}
\begin{figure}[h]
\centering
 \includegraphics[scale=0.35]{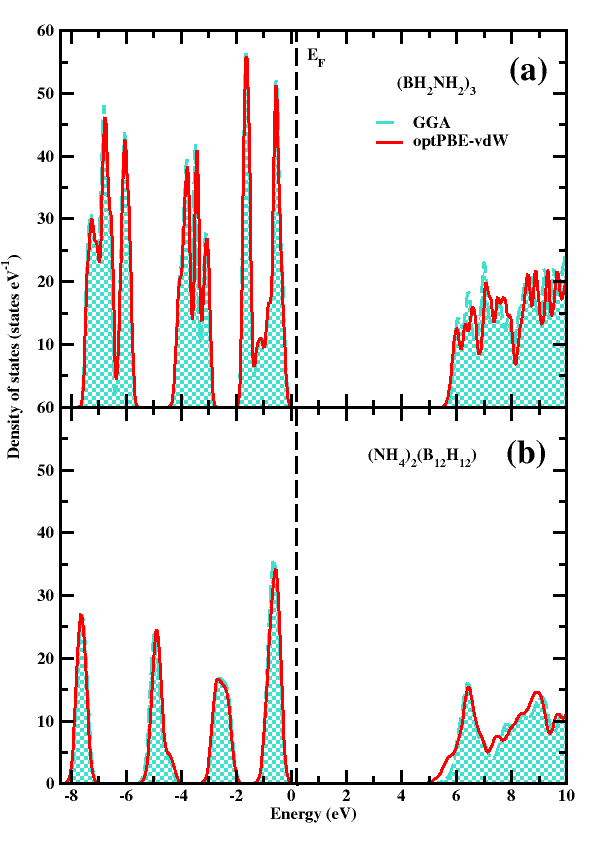}
 \caption{The total density of states (TDOS) obtained from calculations based on GGA and optPBE$-$vdW functionals for (BH$_2$NH$_2$)$_3$ and (NH$_4$)$_2$(B$_{12}$H$_{12}$) are given in (a) and (b), respectively. The solid (red) and dashed (cyan)curves represent the TDOS obtained from optPBE-vdW and GGA calculations.}
 \label{fig:4}
\end{figure}

In order to understand the electronic structure and chemical bonding present between the constituents in (BH$_2$NH$_2$)$_3$ and (NH$_4$)$_2$(B$_{12}$H$_{12}$), the total density of states (TDOS) obtained from GGA and optPBE$-$vdW for (BH$_2$NH$_2$)$_3$ and (NH$_4$)$_2$(B$_{12}$H$_{12}$) are given in Fig. \ref{fig:4} (a) and (b), respectively. From these figures, it is clear that both these compounds posses insulating behavior as there is wide band gap present between valence band maximum (VBM) and conduction band minimum (CBM). For (BH$_2$NH$_2$)$_3$ the band gap value estimated from GGA calculation is 5.55\,eV and that estimated from optPBE$-$vdW is 5.44\,eV. Similarly, for (NH$_4$)$_2$(B$_{12}$H$_{12}$) the band gap value estimated from GGA and optPBE$-$vdW are 5.06\,eV and 4.74\,eV, respectively. From DOS analysis we found that the band gap obtained from GGA is always higher than that obtained from optPBE$-$vdW. If we compare the TDOS distribution for (BH$_2$NH$_2$)$_3$ and (NH$_4$)$_2$(B$_{12}$H$_{12}$), we found that (NH$_4$)$_2$(B$_{12}$H$_{12}$) has more molecular$-$like narrow band behavior. In the case of (BH$_2$NH$_2$)$_3$ the VB comprise of three distinct DOS distributions centered around $-$7\,eV, $-$3\,eV, and $-$1\,eV. These VB states are mainly originating from hybridized bands of N$-$\textit{p}, H$-$\textit{s}, and B$-$\textit{p} states. In the case of (NH$_4$)$_2$(B$_{12}$H$_{12}$), the VB contain four distinct DOS distributions centered around $-$8.5\,eV, $-$5\,ev, $-$2.5\,eV, and $-$0.5\,eV energies. The lowest energy peak at $-$8.5\,eV is originating from the hybridized bands of N$-$\textit{p}, H(+1)$-$\textit{s} states. The VB states at $-$5\,eV is contributed by \textit{p}$-$states of both B and N and the \textit{s}$-$states of both the hydrogen. So, the VB is contributed by all the four constituents. The DOS distribution around $-$2.5\,eV and $-$0.5\,eV are mainly originated from B$-$\textit{p} and H($-$1)$-$\textit{s} electrons. Further, the CBM obtained from optPBE$-$vdW is shifted to about 0.11\,eV and 0.32\,eV for (BH$_2$NH$_2$)$_3$ and (NH$_4$)$_2$(B$_{12}$H$_{12}$), respectively with respective to that from GGA towards lower energy. However, the overall DOS distribution obtained from GGA and optPBE$-$vdW are consistent with each other for both (BH$_2$NH$_2$)$_3$ and (NH$_4$)$_2$(B$_{12}$H$_{12}$) indicate that the intra$-$molecular interaction are not significantly changed by optPBE$-$vdw functional. If we analyze the DOS distribution in the CB of both these compounds, we have found that the molecular$-$like DOS features are disappeared and the DOS distribution is similar to that of 3D inorganic crystals.

\begin{figure}[h]
\centering
 \includegraphics[scale=0.20]{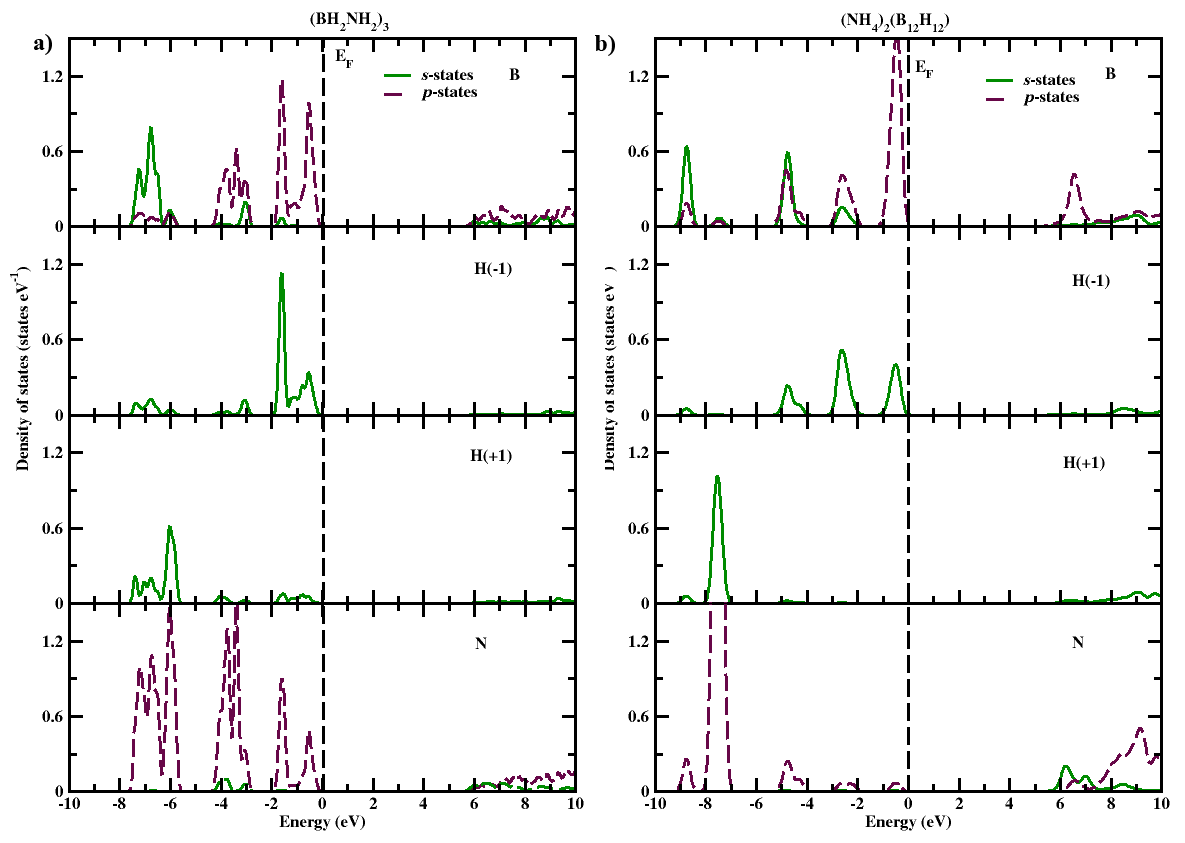}
 \caption{The partial DOS for (BH$_2$NH$_2$)$_3$ and (NH$_4$)$_2$(B$_{12}$H$_{12}$) in (a) and (b), respectively obtained from optPBE$-$vdW functional at the equilibrium volume.}
 \label{fig:5}
\end{figure}

Let us now analyze the partial density of states (PDOS) for (BH$_2$NH$_2$)$_3$ and (NH$_4$)$_2$(B$_{12}$H$_{12}$) obtained from optPBE$-$vdW in Fig. \ref{fig:5} (a) and (b), respectively. The DOS distribution of B and H($-$1) show that the B$-$\textit{p} and H($-$1)$-$\textit{s} states are energetically degenerate from $-$2\,eV to VBM indicating the presence of strong covalent bond. Similarly, for the H($+$1) neighboring to N the \textit{s} electrons are get well localized around $-$5.5\,eV to $-$7.5\,eV and they are energetically degenerate with N$-$\textit{p} states in this energy range indicating that there is noticeable covalent interaction present between them. Because of this covalent hybridization, bonding states are present in this energy range in the COHP of N$-$H($+$1) pair. It may be noted that the electronic states closer to VBM will contribute to the covalent bonding stronger than the electrons in the lower energy states. So, we conclude that the covalent bonding interaction between B$-$H($-$1) is stronger than that between N$-$H($+$1). This observation is consistent with the Mulliken overlap population analysis made in section 3.3.5. 
 
Comparing the PDOS of H($-$1) and H($+$1) we found that the \textit{s}$-$states of H($+$1) is well localized and present in a narrow energy range ($-$8\,eV to $-$5.8\,eV). Also, the number of electrons in the H($+$1) site is much smaller than that of neutral H atom indicating that this hydrogen is present in positive oxidation state. On the other hand, H($-$1)$-$\textit{s} states are present in the whole VB with a strong peak around $-$2\,eV to VBM. Further, the presence of relatively higher amount of electrons in the H($-$1) sites than the neutral H atom indicating that this hydrogen is in negative oxidation states in consistent with various charge analysis discussed later.

Let us now analyze the PDOS of constituents in (NH$_4$)$_2$(B$_{12}$H$_{12}$). There is a well localized DOS distribution present in both N$-$\textit{p} and H($+$1)$-$\textit{s} DOS and are energetically degenerate in the energy range $-$8\,eV to $-$7\,eV indicating the covalent bonding between these atoms. However, very narrow band features along with very small DOS distribution at the H($+$1) sites indicate the presence of ionicity. So, we can conclude that the bonding interactions between N and H($+$1) are of mixed bonding behavior. It may be noted in Fig. \ref{fig:5} that H($-$1) site has three DOS peaks in the VB region. Interestingly, the boron DOS also shows peaks feature in the same energy range. So, this DOS feature can be considered as originating from B$-$\textit{p}$-$H($-$1)$-$\textit{s} covalent hybrid. However, the sharp feature present in the DOS of B$-$\textit{p} states and H($ - $1)$-$\textit{s} states suggesting noticeable ionicity. If we compare the H($-$1) and H($+$1) DOS, we have found that there is very small states available in the H($+$1) sites and also the electrons present in the H($-$1) sites are more than that of neutral H atom clearly show the amphoteric behavior of hydrogen in these systems. 

It may be noted that the VB DOS for both systems have molecular$-$like sharp features and this suggesting the molecular nature of crystal structure in these compound. Due to this molecular$-$like structure the GGA calculated equilibrium interatomic distance are predicted to be much smaller than one expects for usual solids. This could also explain the importance of van der Waals interaction in these systems to predict the structural properties correctly.

\subsection{Chemical Bonding Analysis}
\subsubsection{Charge Density Analysis}
\begin{figure}[h]
\centering
 \includegraphics[scale=0.20]{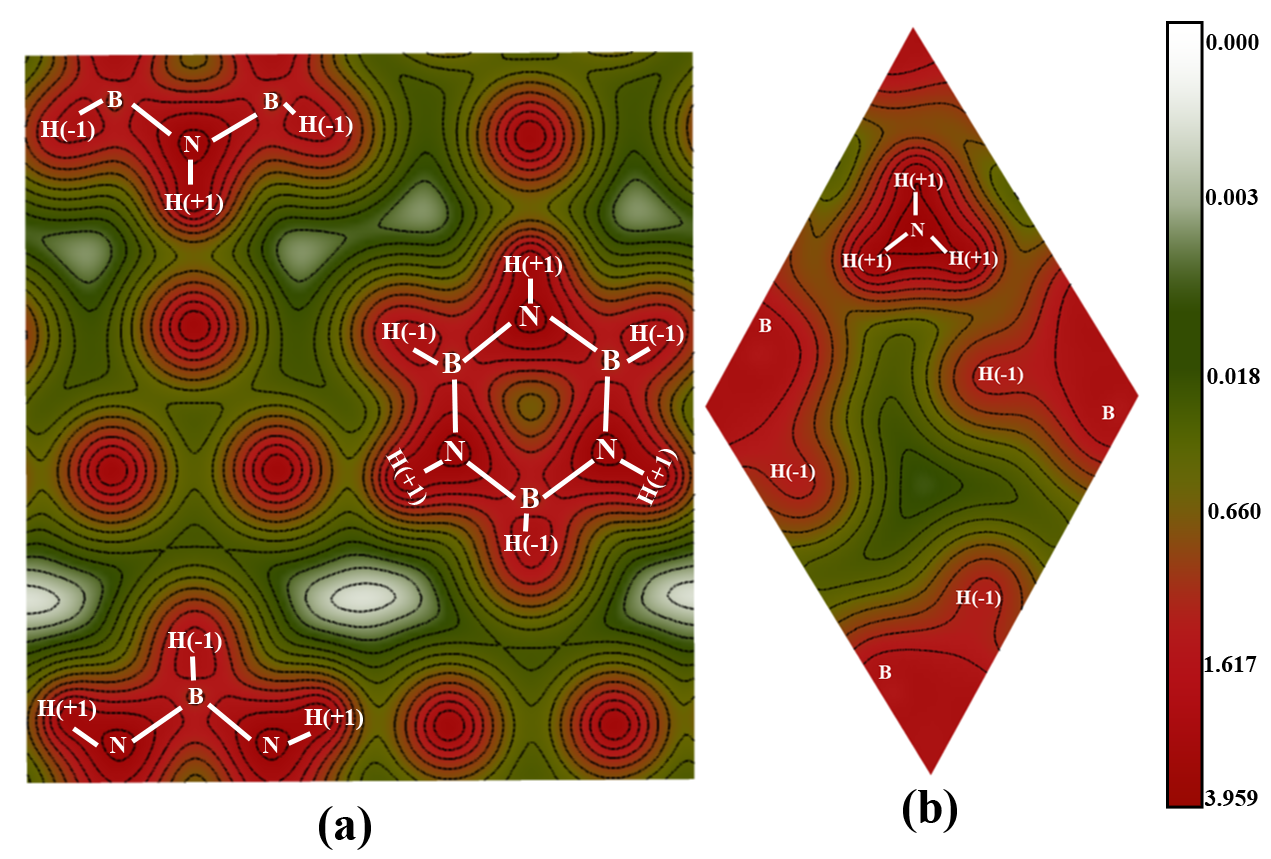}
 \caption{The calculated valance$-$electron density distribution in (a) (BH$_2$NH$_2$)$_3$  and (b) (NH$_4$)$_2$(B$_{12}$H$_{12}$). The electron density (unit of e/\AA $^3$) is projected in a plane where both B$-$H and N$-$H bonds are available.}
 \label{fig:6}
\end{figure}

In order to substantiate the amphoteric behavior of hydrogen and the presence of finite covalent bonding between hydrogen and its neighbors in (BH$_2$NH$_2$)$_3$ and (NH$_4$)$_2$(B$_{12}$H$_{12}$), we have shown charge density distribution for (BH$_2$NH$_2$)$_3$ and (NH$_4$)$_2$(B$_{12}$H$_{12}$) in Fig. \ref{fig:6}(a) and \ref{fig:6} (b), respectively. In general, the spherical distribution of charges around the atomic sites and the negligibly small charge between the constituents describe the ionic bonding.\cite{ravindran2006modeling} In contrast, the anisotropic charge distribution at the atomic sites and finite charge distribution between the constituents is the indication for covalent bonding. The small amount charge density distribution around the B sites and its value is smaller than that of the neutral B atom indicate that the B is in the positive oxidation state. There is a finite charge distribution between hydrogen and its neighbors and its directional dependent behavior indicates the presence of a finite covalent bonding. The noticeable spherical charge distribution at the hydrogen closer to B and its value higher than that of neutral hydrogen atom indicates that the charge transferred from the B is accumulated in these hydrogen sites. However, the finite charge between B and H as well as between H atoms closer to B along with its non$-$spherical distribution indicating the presence of covalent bonding between B$-$H and H$-$H. So, one can conclude that the bonding interaction between B and hydrogen is mixed bonding with iono$-$covalent nature. In order to quantify the covalent and the ionic interaction we have also made various charge analysis schemes and are discussed later.

Let us now analyze the bonding interaction between N and its neighboring hydrogen. As N is more electronegative than the hydrogen, it draws the electron from neighboring hydrogen. Because of this, the charge at the hydrogen sites are smaller than that of neutral atomic hydrogen. However, there is a finite amount of charge present at this hydrogen sites showing that they have not completely donated all their electron to the neighboring N. This indicates the presence of partial ionicity. Further, the N$-$H bond length is very small (1.025~\AA) and this is in consistent with our above observation of positively charged H surrounding the N sites. Our charge density distribution analysis shows that the charge around the N atoms are not spherically distributed and also there is a finite charge exist between N and H indicating the presence of finite covalent bonding between N$-$H. So, the present charge density analysis shows beyond doubt that hydrogen is showing amphoteric behavior in these systems. The presence of partial ionicity in these systems are substantiated by quantifying the charges at various sites using various charge accounting schemes described below.

The B$-$N$-$H form a molecular complex as shown by our charge density distribution plots for both the compounds and hence one can conclude that these compounds can not be considered as three dimensional solids and can be classified as molecular solids. As there is weak inter$-$molecular interaction present within the compound due to the molecular-like solid behavior, our GGA calculations unable to predict reliably their equilibrium structural parameters and hence we have used various van der Waals functionals to predict their structural parameters reliably.

\subsubsection{Electron Localization Function Analysis}
\begin{figure}[h]
\centering
 \includegraphics[scale=0.20]{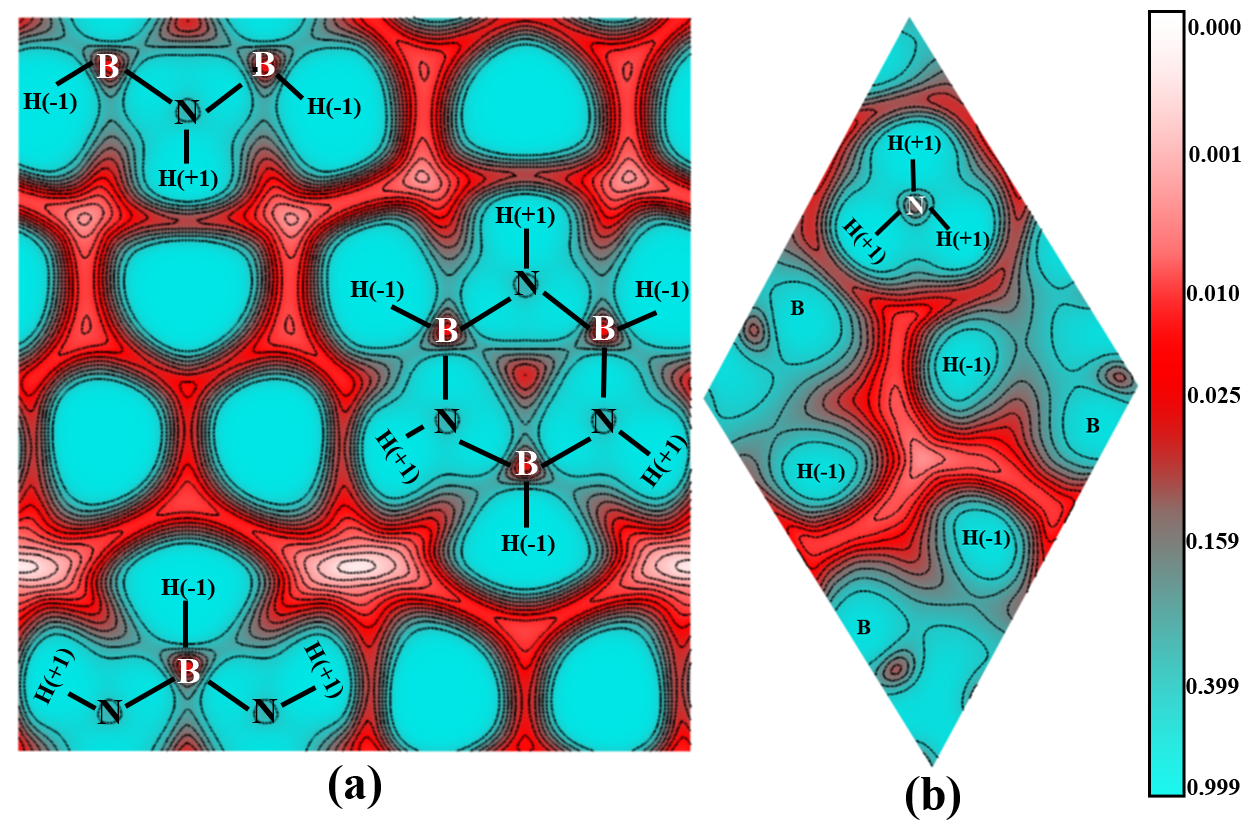}
 \caption{The calculated electron localization function (ELF) for (BH$_{2}$NH$_{2}$)$_{3}$ and (NH$_4$)$_2$(B$_{12}$H$_{12}$) are shown in (a) and (b), respectively obtained from optPBE$-$vdW functional calculation. The ELF distribution is shown for the plane where the B$-$H and N$-$H bonds are present.}
 \label{fig:7}
\end{figure}

\begin{figure}[h]
\centering
 \includegraphics[scale=0.20]{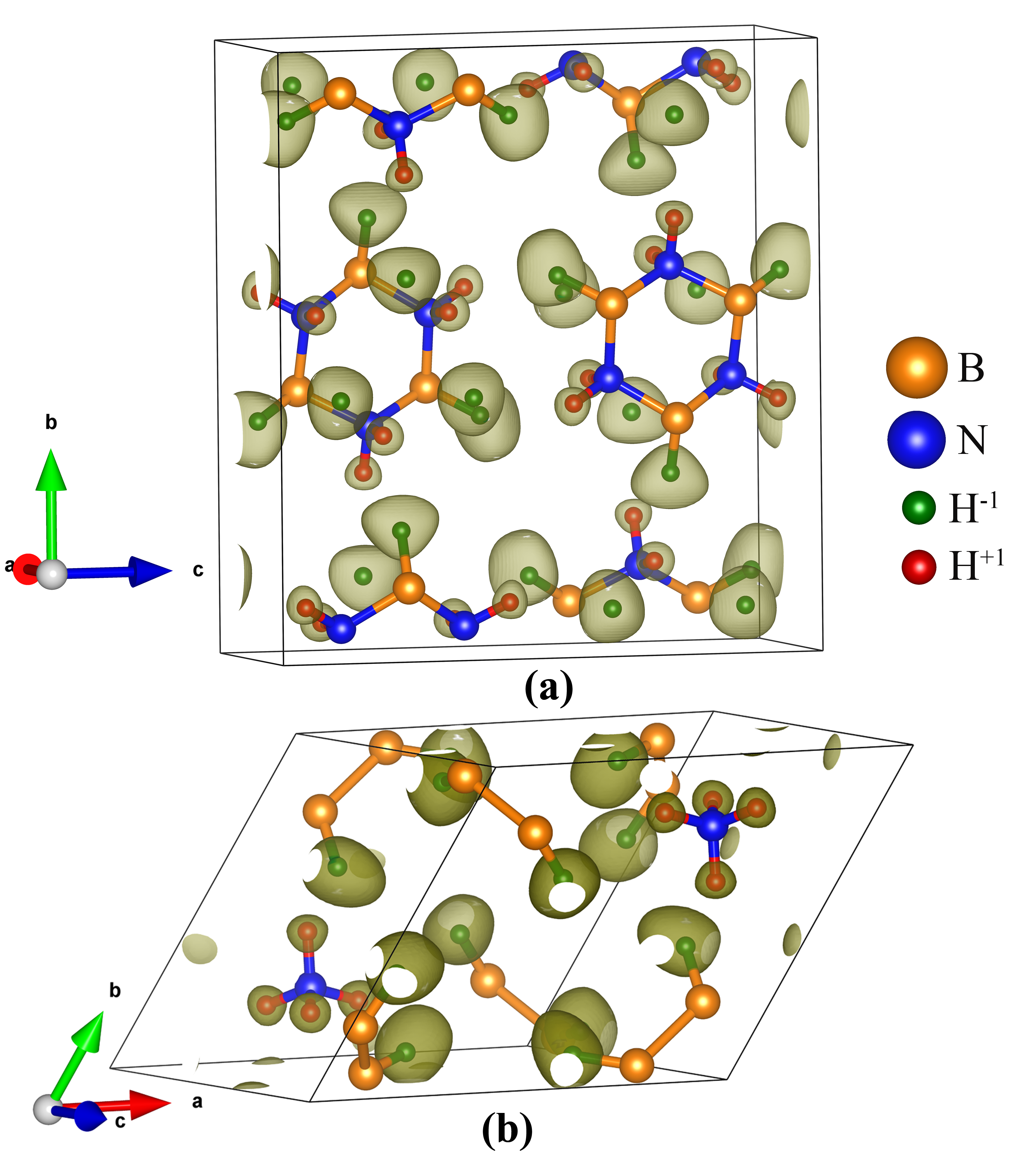}
 \caption{Isosurface(value of 0.97) of the valance$-$electron localization function for (BH$_2$NH$_2$)$_3$ and (NH$_4$)$_2$(B$_{12}$H$_{12}$) are shown in (a) and (b) respectively.} 
 \label{fig:8}
\end{figure}

Electron localization function (ELF) is a property which quantitatively discriminates different kinds of bonding environments. Generally, in a many$-$electron system, the value of ELF becomes close to 1 in the region where electrons are paired to form a covalent bond and also close to 1 for the region with an unpaired electron of a dangling bond ~\cite{vajeeston2004design}. Fig. \ref{fig:7} shows electron localization function for (BH$_2$NH$_2$)$_3$ and (NH$_4$)$_2$(B$_{12}$H$_{12}$) to demonstrate the amphoteric behavior of H in these systems. The ELF value in-between N and H($+$1) in both the systems are very closer to 1, which implicates the covalent type of interaction between  N and H($+$1). On the other hand, the ELF distribution between B and  H($-$1) is  relatively small suggesting that there is an iono$-$covalent bond present between B and H(-1). Moreover, due to the presence of weak inter molecular$-$like bonding interaction between B$-$H$-$N structural sub$-$units, there is negligibly small ELF value present between them. In order to have a clear picture on amphoteric behavior of hydrogen present in these systems, we have plotted the ELF isosurface for the value of 0.97 for (BH$_2$NH$_2$)$_3$ and (NH$_4$)$_2$(B$_{12}$H$_{12}$) in Fig. \ref{fig:8}(a) and \ref{fig:8}(b), respectively. From this figure it is clear that the ELF distribution for hydrogen closer to N have smaller isosurface value than that closer to B. So, the three dimensional visualization of ELF isosurface show beyond doubt the presence of hydrogen with amphoteric behavior in these systems. 

\subsubsection{Born Effective Charge Analysis}
In order to have the deeper insight into the chemical bonding between constituents and the amphoteric behavior of hydrogen, we have calculated the Born effective charge (BEC) and the values are tabulated in Table \ref{tbl:4}. To calculate the Born effective charge tensor we have used the Berry phase approach of the "Modern theory of polarization".\cite{king1993theory} Let us first analyze the calculated BEC of constituents in (BH$_{2}$NH$_{2}$)$_{3}$. As expected from the electronegativity difference between the constituents, there is a substantial ionicity present in these systems. So, the B is in the positively charged state and the H atoms surrounded by the B are negatively charged state. Compared with H, the N atoms are more electronegative and as a consequence of that they draw charge from the H atoms making the H atoms around N are in a positively charged state. However, there is substantial covalent bonding present between the constituents in this system introduce partial ionicity. So, the average diagonal components of BEC always shows a smaller value than the nominal ionic charge for all the constituents. If this system is having pure ionic bonding, one should except that B, N, H($-$1), and H(+1) will have the BEC value of +3, $-$3, $-$1 and +1, respectively. However, due to the partial ionicity, the average diagonal components of BEC at B, N, H($-$1), and H($+$1) are 0.95, $-$1.1, $-$1.05, and 0.26, respectively. If it is a pure ionic system, the off$-$diagonal elements in the BEC tensor are excepted to be zero. But, due to the presence of covalency, the off$-$diagonal elements in the BEC tensor has finite values in all these constituents given in Table \ref{tbl:4}. 

\begin{table*}
\small 
\caption{The calculated values in the Born effective charge tensor (Z$^\ast$) for the constituents of (BH$_2$NH$_2$)$_3$ and (NH$_4$)$_2$(B$_{12}$H$_{12}$)obtained using the GGA and optPBE$-$vdW functional.}
\label{tbl:4}
\begin{tabular*}{\textwidth}{@{\extracolsep{\fill}}cccccccccccc}
\hline
 & Functional & Average of & \multicolumn{9}{c}{$Z^*$(e)}\\ [2.5ex]
 & & the diagonal tensor & $Z_{xx}$ & $Z_{yy}$ & $Z_{zz}$ & $Z_{xy}$ & $Z_{yz}$ & $Z_{zx}$ & $Z_{xz}$ & $Z_{zy}$ & $Z_{yx}$\\
\hline
(BH$_2$NH$_2$)$_3$ & & & & & & & & & & & \\
\hline
B & GGA & 0.950 & 0.768 & 0.958 & 1.126 & $-$0.029 & 0.000 & 0.000 & 0.000 & 0.000 & $ - $0.090\\ [2.5ex]
B & vdW & 1.004 & 0.823 & 1.024 & 1.165 & $-$0.031 & 0.000 & 0.000 & 0.000 & 0.000 & $ - $0.094 \\ [2.5ex]
H($-$1) &GGA & $ - $0.349 & $-$0.581 & $-$0.241 & $-$0.225 & $-$0.033 & 0.000 & 0.000 & 0.000 & 0.000 & 0.080\\ [2.5ex] 
H($-$1)& vdW & $ - $0.362 & $-$0.601 & $-$0.248 & $-$0.239 & $-$0.033 & 0.000 & 0.000 & 0.000 & 0.000 & 0.090\\ [2.5ex]
N & GGA & $ - $0.781 & $-$0.508 & $-$0.949 & $-$0.887 & $-$0.004 & 0.183 & $-$0.053 & 0.026 & 0.166 & $-$0.025\\ [2.5ex]
N &vdW & $ - $0.826 & $-$0.524 & $-$1.021 & $-$0.935 & $-$0.001 & 0.199 & $-$0.073 & 0.024 & 0.176 & $-$0.026\\[2.5ex]
H($+$1)& GGA & 0.262 & 0.298 & 0.189 & 0.300 & $-$0.032 & 0.018 & $-$0.001 & $-$0.041 & 0.032 & $-$0.011\\ [2.5ex]
H($+$1)& vdW & 0.254 & 0.265 & 0.193 & 0.304 & $-$0.031 & 0.015 & $-$0.005 & $-$0.031 & 0.029 & $-$0.009\\[2.5ex]
\hline
(NH$_4$)$_2$(B$_{12}$H$_{12}$) & & & & & & & & & & & \\
\hline
B & GGA & 0.108 & $-$0.123 & 0.178 & 0.023 & 0.000 & 0.105 & 0.000 & 0.000 & 0.184 & 0.000\\[2.5ex]
B & vdW & 0.104 & $-$0.159 & 0.149 & $-$0.006 & 0.000 & 0.136 & 0.000 & 0.000 & 0.188 & 0.000\\[2.5ex]
H($ - $1)& GGA  & $ - $0.232 & $ - $0.119 & $-$0.372 & $-$0.207 & 0.000 & $-$0.168 & 0.000 & 0.000 & $- $0.211 & 0.000\\ [2.5ex]
H($-$1)& vdW & $ - $0.248 & $-$0.123 & $-$0.401 & $-$0.222 & 0.000 & $-$0.165 & 0.000 & 0.000 & $- $0.220 & 0.000\\ [2.5ex]
N& GGA & $ - $0.443 & $-$0.443 & $-$0.443 & $-$0.443 & 0.000 & 0.000 & 0.000 & 0.000 & 0.000 & 0.000 \\ [2.5ex]
N& vdW & $ - $0.670 & $-$0.670& $ - $0.670 & $ - $0.670 & 0.000 & 0.000 & 0.000 & 0.000 & 0.000 & 0.000\\ [2.5ex]
H($+$1) & GGA & 0.422 & 0.422 & 0.422 & 0.422 & 0.151 & 0.151 & 0.151 & 0.151 & 0.151 & 0.151\\ [2.5ex]
H($+$1)& vdW & 0.475 & 0.475 & 0.475 & 0.475 & 0.255 & 0.255 & 0.255 & 0.258 & 0.258 & 0.258\\ [2.5ex]
\hline
\end{tabular*}
\end{table*}
Let us turn our attention towards (NH$_4$)$_2$(B$_{12}$H$_{12}$) to analyze the charge state of the constituents based on the calculated BEC. In this compound, each B is surrounded by five B and one H as neighbors. The charge density and DOS analyses show that there is strong covalent bonding present between  B with the neighboring B atoms. So, there is limited charge only available at the B site to donate to the neighboring H atoms and hence the  calculated BEC at the H($-$1) site is smaller ($-$0.698) than the nominal ionic value of $-$1. Further, the anisotropic value of diagonal components of BEC at both B and H($-$1) sites along with the finite value of off$-$diagonal elements of BEC indicate the covalent bonding between B and H($-$1). Due to relatively high electronegativity of N, it draws more charge from the neighboring H sites make these H into the protonic state. However, owing to the partial covalency, the H neighboring to N are having a smaller average value of BEC than the nominal ionic charge of $+$1. So, the calculated BEC values confirm beyond doubt that the H atoms in (BH$_{2}$NH$_{2}$)$_{3}$ and (NH$_4$)$_2$(B$_{12}$H$_{12}$) are in an amphoteric state with H atoms neighboring to B are in positive oxidation state, whereas, that neighboring to N are in negative oxidation state.
\subsubsection{COHP Analysis}
\begin{figure}[h]
\centering
 \includegraphics[scale=0.20]{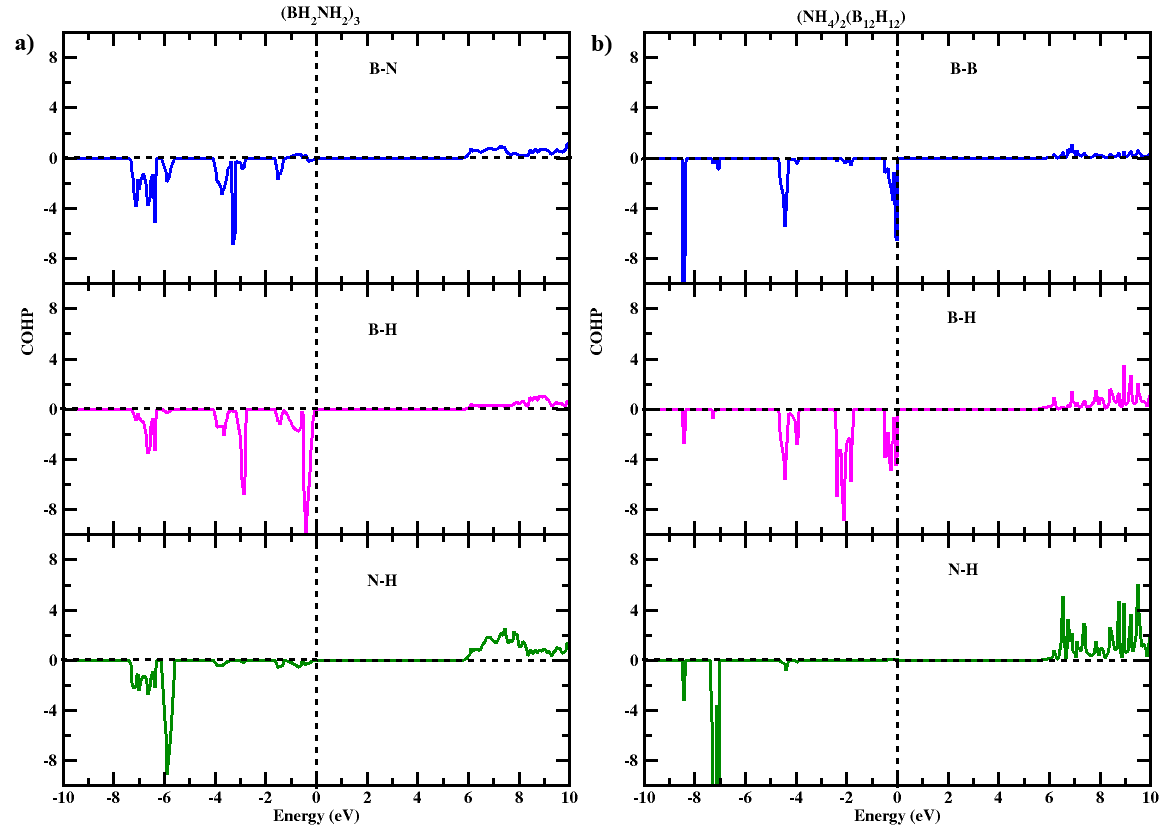}
 \caption{COHP between constituents in (BH$_2$NH$_2$)$_3$ and (NH$_4$)$_2$(B$_{12}$H$_{12}$) are shown in (a) and (b), respectively obtained using optPBE$-$vdW functionals.}
 \label{fig:9}
\end{figure}

In order to evaluate the bond strength between the involved atoms, we have performed COHP analysis. The COHP interaction between the involved atoms in (BH$_{2}$NH$_{2}$)$_{3}$ and (NH$_4$)$_2$(B$_{12}$H$_{12}$) are shown in Fig. \ref{fig:9} (a) and (b), respectively. A negative value of COHP indicates bonding states and a positive value of COHP will show anti$-$bonding states. The band filling of the bonding states analyses show that one can obtain maximum stability when all the bonding states are filled and the anti$-$bonding states are empty. In the present study also, both the systems, all the bonding states are present within the VB and hence they are filled and the anti$-$bonding states are located in the CB and are empty making these systems stable.

If we analyse the calculated COHP between the constituents in (BH$_{2}$NH$_{2}$)$_{3}$ we found that the bonding states of N$-$H(+1) pair are located from $- $8~eV to $-$5.8~eV and the anti$-$bonding states are located above 5.8~eV. There is a noticeable covalent bonding present in the N$-$H($+$1) pair and hence the bonding states are well localized around $-$5.8~eV. The observation of covalent bonding behavior from COHP analysis is in consistent with that from DOS analysis. On the other hand, for the B$-$H($-$1) pair, the bonding states are located in the entire VB region with three distinct peaks at $-$5.4~~eV, $-$3.3~eV, and $-$0.7~eV  and the antibonding states are located above 6~eV. The integarated COHP (ICOHP) value upto Fermi level will give the bond strength of the bonding pair and hence we have calculated the ICOHP value between the constituents. The calculated ICOHP value for B$-$H($-$1) is $-$6.86\,eV and that for N$-$H($+$1) is $-$8.16\,eV. This indicates that the H($+$1) ions are strongly bonded with the neighboring N than H($-$1) with B and hence it is expected that the H($-$1) attached to the B site can desorb relatively easily. Though the PDOS, charge density, ELF and Mulliken population analyses show that the covalent interaction between N and H ($+$1) is weaker than that between B and H($-$1), the ionic interaction between N$-$H ($+$1) is stronger than that between B$-$H($-$1) (please also see other chemical bonding analyses). The complementary interaction from the ionic as well as covalent bonding to the bond strength make N$-$H($+$1) bond stronger than the B$-$H ($-$1) bond as our analysis suggest.

The PDOS for N in (BH$_{2}$NH$_{2}$)$_{3}$ show (Fig. \ref{fig:5}) that there is a broad peak in the lowest energy region of VB between $-$7.5~eV to $-$5.7~eV and it may be noted that the N$-$H(+1) bonding states are present around $-$6 eV. The states below $-$6~eV are dominated by non$-$bonding states. Above this DOS distribution in VB, there are finite DOS distributed around $-$3.8~eV, $-$1.8~eV, and $-$0.3~eV. Among these DOS peaks, one around $-$3.8~eV and $-$1.8~eV are contributing to B$-$N bonding hybrid, whereas, the DOS around $-$0.3~eV are mainly having non$-$bonding character. The B$-$\textit{p} DOS contribute at the topmost region of VB is located between $-$2~eV to VBM with two distinct peaks at $-$1.8~eV and $-$0.5~eV. Among these two DOS peaks, one near the VBM is mainly contributed by B$-$H bonding hybrid and the DOS around $-$1.8~eV is dominated by non$-$bonding B$-$2\textit{p} electrons. The calculated ICOHP value for B$-$N pair is $-$7.97~eV and this indicates that its bond strength is in between that of N$-$H and B$-$H bond.

The COHP analysis of (NH$_4$)$_2$(B$_{12}$H$_{12}$) reveals that the bonding states of N$-$H($+$1) pair is present in a very narrow energy range around $-$7~eV and the anti$-$bonding states are located above 5.4~eV. This suggests that there is considerable covalent bonding present between N and H($+$1). However, the bonding states for B$-$H($-$1) pair is located in a broad energy range from $-$5~eV to VBM and the anti$-$bonding states are located above 5.4~eV. In this compound there is a substantial number of B$-$B pairs present and the COHP of B$-$B pair has three distinct bonding states located at $-$8.5~eV, $-$4.5~eV, and $-$0.5~eV and the anti$-$bonding states are present above 5.4~eV. Apart from these three energy states, there is finite DOS distributed around $-$2.5~eV and is contributing to B$-$H($-$1) bonding hybrid. The calculated ICOHP values for N$-$H($+$1) and B$-$H($-$1) bonding hybrids in this compound are $-$8.52~eV and $-$6.52~eV, respectively indicating that H is relatively weakly bonded with boron than with N. So, one would expect that during the hydrogen decomposition process hydrogen neighboring to B will release first. The inter$-$atomic distance between B and N in this compound is 2.93 \AA~and hence there is no direct bonding pairs form between them. The calculated ICOHP value for B$-$B pair is $-$4.44~eV and this indicate that the bonding interaction between neighboring B atoms are weaker than that between B$-$H as well as N$-$H.

\subsubsection{Bader and Mulliken Charge Analyses}
In order to quantify the bonding and estimate the electron distribution at the participating atoms, we have calculated Bader charges (BC) using Bader's atoms$-$in$-$molecule approach. Bader's theory of atoms$-$in$-$molecule is often useful for charge analysis.\cite{bader1990atoms, henkelman2006fast} The theory provides a definition for chemical bonding that gives numerical values for bond strength. The zero flux surfaces in the charge density are used to define each basin belonging to a particular atom. The charge assigned to this atom is then obtained by integrating the charge density over the whole basin. The BC of participating atoms in (BH$_{2}$NH$_{2}$)$_{3}$ and (NH$_{4}$)$_{2}$(B$_{12}$H$_{12}$) are calculated and tabulated in Table \ref{tbl:5}.\\

In (BH$_{2}$NH$_{2}$)$_{3}$, the BC values for B and H($-$1) are $+$1.85 e and $-$0.55 e, respectively. This suggests that $+$1.85~e are transferred from B to H($-$1). On the other hand, the BC value for N and H($+$1) are $-$1.60 e and $+$0.454~e, respectively which indicates the presence of partial ionicity between N and H($+$1). The charge density distribution analysis also suggests that there is a noticeable covalent character present between B and N, and hence consistent with BC analysis. From the Table \ref{tbl:5}, it is clear that N has the total charge of 4.0 electrons, in which 1.6 electrons are shared with its neighboring hydrogen atom and the remaining electrons (i.e. for about 2.4~e) are shared with neighboring B atoms in the system. The H($-$1) and H($+$1) BC values in (BH$_{2}$ NH$_{2}$)$_{3}$ clearly shows the presence of amphoteric hydrogen in the system where H($+$1) donated $+$0.461 electrons to the host lattice and  H($-$1) accepted $-$0.583 electrons from the neighboring constituents. \\ 

\begin{table}[h]
\small
\caption{The calculated Bader effective charge (BC in e), Mulliken effective charge (MC in e) at various sites in (BH$_{2}$ NH$_{2}$)$_{3}$ and (NH$_4$)$_2$(B$_{12}$H$_{12}$).}
\label{tbl:5}
\begin{tabular*}{0.48\textwidth}{@{\extracolsep{\fill}}llll} 
\hline
 Compound & Atom site & {BC} & MC \\
\hline
 (BH$_{2}$ NH$_{2}$)$_{3}$ & B & 1.960 & 0.19 \\[1.5ex]
 & H($-$1) & $-$0.583 & $-$0.09 \\ [1.5ex] 
 & N & $-$1.635 & $-$0.80 \\ [1.5ex]
 & H($+$1) & 0.461 & 0.38 \\ [1.5ex]
\hline
 (NH$_4$)$_2$(B$_{12}$H$_{12}$) & B & 0.451 & 0.19\\ [1.5ex]
 & H($-$1) & $-$0.581 & $-$0.03\\ [1.5ex]
 & N & $-$1.311 & $-$0.76\\ [1.5ex]
 & H($+$1) & 0.525 & 0.44\\[1.5ex]
\hline
\end{tabular*}
\end{table}

In (NH$_{4}$)$_{2}$(B$_{12}$ H$_{12}$), the BC of B and H($-$1) reveals the partial ionic interaction between B and H($-$1), because, the boron has transferred $+$0.451 electrons only to its neighboring hydrogen (H($-$1)). Moreover, there is covalent interaction present between B and B, which we have already discussed in the previous sections such as structural analysis, COHP analysis, etc. Hence, it is evident that around 1.6 electrons are shared by B with the neighboring B atoms in this system. The BC value for N and H($+$1) sites are $-$1.205 and $+$0.525 electrons, respectively. Here, the H$+$1 attached to the nitrogen atom donated $+$0.525 electrons and the H($-$1) attached to the boron accepted $-$0.581 electrons and hence this confirms the presence of amphoteric hydogen in this compound.~\cite{ramzan2009structural,hamilton2009b,matus2009fundamental}\\

The Mulliken charges (MC) calculated based on Mulliken population analysis \cite{mulliken1955rs} for (BH$_{2}$ NH$_{2}$)$_{3}$ and (NH$_4$)$_2$(B$_{12}$H$_{12}$) are listed in Table \ref{tbl:3}. The overlap population between atoms give information about the covalent bonding interaction. Let us first analyze the overlap population between H with its neighbors in (BH$_{2}$ NH$_{2}$)$_{3}$. The calculated overlap population between B$-$H($-$1) and N$-$H($+$1) are 1.03 and 0.70, respectively. In a pure ionic system, the overlap population between the constituents will be negligibly small. However, the noticeable amount of overlap charge presents between H and its neighbors clearly indicating the presence of covalency. The overlap population between B$-$H and N$-$H in (NH$_4$)$_2$(B$_{12}$H$_{12}$) are 1.04 and 0.70, respectively. The present Mulliken population analysis suggests that the B$-$H($-$1) bond is more covalent than the N$-$H($+$1) bond in both these systems. The calculated MC for B and H($-$1) are 0.19 and $-$0.09 electrons, respectively and that for H($+$1) and N are $-$0.80 and 0.38 electrons, respectively in (BH$_{2}$ NH$_{2}$)$_{3}$. In (NH$_4$)$_2$(B$_{12}$H$_{12}$) also, the MC at B and H($-$1) are 0.19 and $-$0.09 electrons and that at N and H(+1) are $-$0.76 and 0.44 electrons, respectively. Though the calculated MC values are smaller than the corresponding BC values, the Mulliken population analysis also shows the amphoteric behavior of hydrogen in these compounds. 

\section{Summary}
Using \textit{ab$-$initio} band structure method we have analyzed the chemical bonding between the constituents in (BH$_{2}$NH$_{2}$)$_{3}$ and (NH$_4$)$_2$(B$_{12}$H$_{12}$).

$\bullet$ We have observed that the total energy vs volume curve obtained from GGA calculation is (i.e.~without vdW correction) overestimated with respect to experimental value by about 4.40\%, and this indicates that the GGA functional unable to account the interaction between the molecular$-$like structural sub$-$units present in these systems. 

$\bullet$ After considering 12 different van der Waals functionals into the total energy calculations we have found that rPW86$-$vdW and optPBE$-$vdW are more appropriate to predict structural properties of molecular$-$like hydrides such as (BH$_2$NH$_2$)$_3$ and (NH$_4$)$_2$(B$_{12}$H$_{12}$). However, optPBE$-$vdW yield overall good agreement with experimental equilibrium volume than other functionals considered in the present study for both the systems and hence results obtained based on optPBE$-$vdW was used for further analysis.

$\bullet$ The DOS analysis clearly shows that (BH$_{2}$NH$_{2}$)$_{3}$ and (NH$_4$)$_2$(B$_{12}$H$_{12}$) have insulating behavior with the band gap value of 5.44\,eV and 4.74\,eV, respectively. We found that the bonding interaction between N$-$H($+$1) and B$-$H($-$1) are iono$-$covalent character. The partial DOS analyses show the amphoteric behavior of hydrogen in both the systems that hydrogen closer to B has negative oxidation state and that closer to N has the positive oxidation state.  

$\bullet$ The charge density distribution and ELF analyses clearly show that the bonding in B$-$H($-$1) and N$-$H($+$1) are mixed iono$-$covalent character. From these analyses, we confirm the presence of weak inter$-$molecular interactions in these systems. The 3 dimensional visualization of ELF isosurface clearly demonstrates the amphoteric behavior of hydrogen in these systems.

$\bullet$ The BEC analyses for (BH$_{2}$NH$_{2}$)$_{3}$ and (NH$_4$)$_2$(B$_{12}$H$_{12}$) show the presence of substantial covalent bonding and partial ionicity in the bonding between constituents. The calculated BEC from modern theory of polarization confirm that H atoms in (BH$_{2}$NH$_{2}$)$_{3}$ and (NH$_4$)$_2$(B$_{12}$H$_{12}$) are in amphoteric state with H neighboring to B are in positive oxidation state whereas that neighboring to N are in negative oxidation state.

$\bullet$ The COHP analysis between constituents in (BH$_{2}$NH$_{2}$)$_{3}$ and (NH$_4$)$_2$(B$_{12}$H$_{12}$) shows that there is strong bonding interaction between the B$-$H and N$-$H pairs, compare to other constituents pairs. From the calculated ICOHP values  for various bonding pairs we have found that hydrogen is strongly bonded with N than with B and hence we conclude that during hydrogen desorption process the negatively charged hydrogen will be released first in both the systems.

$\bullet$The Bader and the Mulliken charge analyses, reveal that there is a presence of noticeable covalency and partial ionicity in the bonds between the constituents in both (BH$_{2}$NH$_{2}$)$_{3}$ and (NH$_4$)$_2$(B$_{12}$H$_{12}$) systems. The BC and MC values clearly show the presence of amphoteric hydrogen in these compounds.

As we demonstrated here, it is possible to have hydrogen storage systems where one can keep positively and negatively charged hydrogen within the same structural framework. If one can keep positively and negatively charged hydrogen ions adjacent to each other they can be placed much closer due to Coulombic attraction between them than for example hydrogen in metals and hence improves the volumetric hydrogen storage capacity. We hope that the present study will motivate further research in this direction by identifying practical hydrogen storage materials those possessing hydrogen with amphoteric behavior and thus high volumetric capacity.

\section*{Acknowledgment}
The authors are grateful to the Department of Science and Technology, India for the funding support via Grant No. SR/NM/NS$-$1123/2013 and the  Research Council of Norway for providing computer time (under the project number NN2875k) at the Norwegian supercomputer facility.


\balance


\bibliography{references} 

\providecommand*{\mcitethebibliography}{\thebibliography}
\csname @ifundefined\endcsname{endmcitethebibliography}
{\let\endmcitethebibliography\endthebibliography}{}
\begin{mcitethebibliography}{82}
\providecommand*{\natexlab}[1]{#1}
\providecommand*{\mciteSetBstSublistMode}[1]{}
\providecommand*{\mciteSetBstMaxWidthForm}[2]{}
\providecommand*{\mciteBstWouldAddEndPuncttrue}
  {\def\EndOfBibitem{\unskip.}}
\providecommand*{\mciteBstWouldAddEndPunctfalse}
  {\let\EndOfBibitem\relax}
\providecommand*{\mciteSetBstMidEndSepPunct}[3]{}
\providecommand*{\mciteSetBstSublistLabelBeginEnd}[3]{}
\providecommand*{\EndOfBibitem}{}
\mciteSetBstSublistMode{f}
\mciteSetBstMaxWidthForm{subitem}
{(\emph{\alph{mcitesubitemcount}})}
\mciteSetBstSublistLabelBeginEnd{\mcitemaxwidthsubitemform\space}
{\relax}{\relax}

\bibitem[Z{\"u}ttel(2007)]{zuttel2007hydrogen}
A.~Z{\"u}ttel, \emph{MITIG ADAPT STRAT GL}, 2007, \textbf{12}, 343--365\relax
\mciteBstWouldAddEndPuncttrue
\mciteSetBstMidEndSepPunct{\mcitedefaultmidpunct}
{\mcitedefaultendpunct}{\mcitedefaultseppunct}\relax
\EndOfBibitem
\bibitem[Z{\"u}ttel(2003)]{zuttel2003materials}
A.~Z{\"u}ttel, \emph{Mater. Today}, 2003, \textbf{6}, 24--33\relax
\mciteBstWouldAddEndPuncttrue
\mciteSetBstMidEndSepPunct{\mcitedefaultmidpunct}
{\mcitedefaultendpunct}{\mcitedefaultseppunct}\relax
\EndOfBibitem
\bibitem[Jena(2011)]{jena2011materials}
P.~Jena, \emph{J. Phys. Chem. Lett}, 2011, \textbf{2}, 206--211\relax
\mciteBstWouldAddEndPuncttrue
\mciteSetBstMidEndSepPunct{\mcitedefaultmidpunct}
{\mcitedefaultendpunct}{\mcitedefaultseppunct}\relax
\EndOfBibitem
\bibitem[Sakintuna \emph{et~al.}(2007)Sakintuna, Lamari-Darkrim, and
  Hirscher]{sakintuna2007metal}
B.~Sakintuna, F.~Lamari-Darkrim and M.~Hirscher, \emph{Int. J. Hydrog. Energy},
  2007, \textbf{32}, 1121--1140\relax
\mciteBstWouldAddEndPuncttrue
\mciteSetBstMidEndSepPunct{\mcitedefaultmidpunct}
{\mcitedefaultendpunct}{\mcitedefaultseppunct}\relax
\EndOfBibitem
\bibitem[Schlapbach and Z{\"u}ttel(2011)]{schlapbach2011hydrogen}
L.~Schlapbach and A.~Z{\"u}ttel, in \emph{Materials For Sustainable Energy: A
  Collection of Peer-Reviewed Research and Review Articles from Nature
  Publishing Group}, World Scientific, 2011, pp. 265--270\relax
\mciteBstWouldAddEndPuncttrue
\mciteSetBstMidEndSepPunct{\mcitedefaultmidpunct}
{\mcitedefaultendpunct}{\mcitedefaultseppunct}\relax
\EndOfBibitem
\bibitem[Matar(2010)]{matar2010intermetallic}
S.~F. Matar, \emph{Prog. Solid State Chem.}, 2010, \textbf{38}, 1--37\relax
\mciteBstWouldAddEndPuncttrue
\mciteSetBstMidEndSepPunct{\mcitedefaultmidpunct}
{\mcitedefaultendpunct}{\mcitedefaultseppunct}\relax
\EndOfBibitem
\bibitem[Ley \emph{et~al.}(2014)Ley, Jepsen, Lee, Cho, Von~Colbe, Dornheim,
  Rokni, Jensen, Sloth, Filinchuk,\emph{et~al.}]{ley2014complex}
M.~B. Ley, L.~H. Jepsen, Y.-S. Lee, Y.~W. Cho, J.~M.~B. Von~Colbe, M.~Dornheim,
  M.~Rokni, J.~O. Jensen, M.~Sloth, Y.~Filinchuk \emph{et~al.}, \emph{Mater.
  Today}, 2014, \textbf{17}, 122--128\relax
\mciteBstWouldAddEndPuncttrue
\mciteSetBstMidEndSepPunct{\mcitedefaultmidpunct}
{\mcitedefaultendpunct}{\mcitedefaultseppunct}\relax
\EndOfBibitem
\bibitem[Yu \emph{et~al.}(2017)Yu, Tang, Sun, Ouyang, and Zhu]{yu2017recent}
X.~Yu, Z.~Tang, D.~Sun, L.~Ouyang and M.~Zhu, \emph{Prog. Mater. Sci.}, 2017,
  \textbf{88}, 1--48\relax
\mciteBstWouldAddEndPuncttrue
\mciteSetBstMidEndSepPunct{\mcitedefaultmidpunct}
{\mcitedefaultendpunct}{\mcitedefaultseppunct}\relax
\EndOfBibitem
\bibitem[Bogdanovi{\'c} and Schwickardi(1997)]{bogdanovic1997ti}
B.~Bogdanovi{\'c} and M.~Schwickardi, \emph{J Alloys Compd.}, 1997,
  \textbf{253}, 1--9\relax
\mciteBstWouldAddEndPuncttrue
\mciteSetBstMidEndSepPunct{\mcitedefaultmidpunct}
{\mcitedefaultendpunct}{\mcitedefaultseppunct}\relax
\EndOfBibitem
\bibitem[Sch{\"u}th \emph{et~al.}(2004)Sch{\"u}th, Bogdanovi{\'c}, and
  Felderhoff]{schuth2004light}
F.~Sch{\"u}th, B.~Bogdanovi{\'c} and M.~Felderhoff, \emph{Chem. Comm}, 2004,
  2249--2258\relax
\mciteBstWouldAddEndPuncttrue
\mciteSetBstMidEndSepPunct{\mcitedefaultmidpunct}
{\mcitedefaultendpunct}{\mcitedefaultseppunct}\relax
\EndOfBibitem
\bibitem[Li \emph{et~al.}(2011)Li, Yan, Orimo, Z{\"u}ttel, and
  Jensen]{li2011recent}
H.-W. Li, Y.~Yan, S.-i. Orimo, A.~Z{\"u}ttel and C.~M. Jensen, \emph{Energies},
  2011, \textbf{4}, 185--214\relax
\mciteBstWouldAddEndPuncttrue
\mciteSetBstMidEndSepPunct{\mcitedefaultmidpunct}
{\mcitedefaultendpunct}{\mcitedefaultseppunct}\relax
\EndOfBibitem
\bibitem[Chen \emph{et~al.}(2002)Chen, Xiong, Luo, Lin, and
  Tan]{chen2002interaction}
P.~Chen, Z.~Xiong, J.~Luo, J.~Lin and K.~L. Tan, \emph{Nature}, 2002,
  \textbf{420}, 302\relax
\mciteBstWouldAddEndPuncttrue
\mciteSetBstMidEndSepPunct{\mcitedefaultmidpunct}
{\mcitedefaultendpunct}{\mcitedefaultseppunct}\relax
\EndOfBibitem
\bibitem[Khan and Jain(2016)]{khan2016chloride}
J.~Khan and I.~Jain, \emph{Int. J. Hydrog. Energy}, 2016, \textbf{41},
  8271--8276\relax
\mciteBstWouldAddEndPuncttrue
\mciteSetBstMidEndSepPunct{\mcitedefaultmidpunct}
{\mcitedefaultendpunct}{\mcitedefaultseppunct}\relax
\EndOfBibitem
\bibitem[Resan \emph{et~al.}(2005)Resan, Hampton, Lomness, and
  Slattery]{resan2005effects}
M.~Resan, M.~D. Hampton, J.~K. Lomness and D.~K. Slattery, \emph{Int. J.
  Hydrog. Energy}, 2005, \textbf{30}, 1413--1416\relax
\mciteBstWouldAddEndPuncttrue
\mciteSetBstMidEndSepPunct{\mcitedefaultmidpunct}
{\mcitedefaultendpunct}{\mcitedefaultseppunct}\relax
\EndOfBibitem
\bibitem[Kim \emph{et~al.}(2008)Kim, Jin, Shim, and Cho]{kim2008reversible}
J.-H. Kim, S.-A. Jin, J.-H. Shim and Y.~W. Cho, \emph{Scr. Mater}, 2008,
  \textbf{58}, 481--483\relax
\mciteBstWouldAddEndPuncttrue
\mciteSetBstMidEndSepPunct{\mcitedefaultmidpunct}
{\mcitedefaultendpunct}{\mcitedefaultseppunct}\relax
\EndOfBibitem
\bibitem[Rude \emph{et~al.}(2011)Rude, Nielsen, Ravnsbaek, B{\"o}senberg, Ley,
  Richter, Arnbjerg, Dornheim, Filinchuk,
  Besenbacher,\emph{et~al.}]{rude2011tailoring}
L.~H. Rude, T.~K. Nielsen, D.~B. Ravnsbaek, U.~B{\"o}senberg, M.~B. Ley,
  B.~Richter, L.~M. Arnbjerg, M.~Dornheim, Y.~Filinchuk, F.~Besenbacher
  \emph{et~al.}, \emph{Phys. Status Solidi A}, 2011, \textbf{208},
  1754--1773\relax
\mciteBstWouldAddEndPuncttrue
\mciteSetBstMidEndSepPunct{\mcitedefaultmidpunct}
{\mcitedefaultendpunct}{\mcitedefaultseppunct}\relax
\EndOfBibitem
\bibitem[Dafert and Miklauz(1910)]{dafert1910new}
F.~Dafert and R.~Miklauz, \emph{Monatsh. Chem}, 1910, \textbf{31}, 981\relax
\mciteBstWouldAddEndPuncttrue
\mciteSetBstMidEndSepPunct{\mcitedefaultmidpunct}
{\mcitedefaultendpunct}{\mcitedefaultseppunct}\relax
\EndOfBibitem
\bibitem[Ruff and Goerges(1911)]{ruff1911lithium}
O.~Ruff and H.~Goerges, \emph{Ber. Dtsch. Chem. Ges}, 1911, \textbf{44},
  502--506\relax
\mciteBstWouldAddEndPuncttrue
\mciteSetBstMidEndSepPunct{\mcitedefaultmidpunct}
{\mcitedefaultendpunct}{\mcitedefaultseppunct}\relax
\EndOfBibitem
\bibitem[Luo(2004)]{luo2004linh2}
W.~Luo, \emph{J Alloys Compd.}, 2004, \textbf{381}, 284--287\relax
\mciteBstWouldAddEndPuncttrue
\mciteSetBstMidEndSepPunct{\mcitedefaultmidpunct}
{\mcitedefaultendpunct}{\mcitedefaultseppunct}\relax
\EndOfBibitem
\bibitem[Wu(2008)]{wu2008structure}
H.~Wu, \emph{J Am. Chem. Soc}, 2008, \textbf{130}, 6515--6522\relax
\mciteBstWouldAddEndPuncttrue
\mciteSetBstMidEndSepPunct{\mcitedefaultmidpunct}
{\mcitedefaultendpunct}{\mcitedefaultseppunct}\relax
\EndOfBibitem
\bibitem[Klerke \emph{et~al.}(2008)Klerke, Christensen, N{\o}rskov, and
  Vegge]{klerke2008ammonia}
A.~Klerke, C.~H. Christensen, J.~K. N{\o}rskov and T.~Vegge, \emph{J Mater
  Chem}, 2008, \textbf{18}, 2304--2310\relax
\mciteBstWouldAddEndPuncttrue
\mciteSetBstMidEndSepPunct{\mcitedefaultmidpunct}
{\mcitedefaultendpunct}{\mcitedefaultseppunct}\relax
\EndOfBibitem
\bibitem[Corfield and Shore(1973)]{corfield1973crystal}
P.~Corfield and S.~Shore, \emph{J. Am. Chem. Soc}, 1973, \textbf{95},
  1480--1487\relax
\mciteBstWouldAddEndPuncttrue
\mciteSetBstMidEndSepPunct{\mcitedefaultmidpunct}
{\mcitedefaultendpunct}{\mcitedefaultseppunct}\relax
\EndOfBibitem
\bibitem[Yang \emph{et~al.}(2008)Yang, Lamsal, Cai, James, and
  Yelon]{yang2008structural}
J.~Yang, J.~Lamsal, Q.~Cai, W.~James and W.~Yelon, \emph{Appl Phys Lett.},
  2008, \textbf{92}, 091916\relax
\mciteBstWouldAddEndPuncttrue
\mciteSetBstMidEndSepPunct{\mcitedefaultmidpunct}
{\mcitedefaultendpunct}{\mcitedefaultseppunct}\relax
\EndOfBibitem
\bibitem[Dalebrook \emph{et~al.}(2013)Dalebrook, Gan, Grasemann, Moret, and
  Laurenczy]{dalebrook2013hydrogen}
A.~F. Dalebrook, W.~Gan, M.~Grasemann, S.~Moret and G.~Laurenczy, \emph{Chem.
  Comm}, 2013, \textbf{49}, 8735--8751\relax
\mciteBstWouldAddEndPuncttrue
\mciteSetBstMidEndSepPunct{\mcitedefaultmidpunct}
{\mcitedefaultendpunct}{\mcitedefaultseppunct}\relax
\EndOfBibitem
\bibitem[Orimo \emph{et~al.}(2007)Orimo, Nakamori, Eliseo, Z{\"u}ttel, and
  Jensen]{orimo2007complex}
S.-i. Orimo, Y.~Nakamori, J.~R. Eliseo, A.~Z{\"u}ttel and C.~M. Jensen,
  \emph{Chem. Rev.}, 2007, \textbf{107}, 4111--4132\relax
\mciteBstWouldAddEndPuncttrue
\mciteSetBstMidEndSepPunct{\mcitedefaultmidpunct}
{\mcitedefaultendpunct}{\mcitedefaultseppunct}\relax
\EndOfBibitem
\bibitem[Van~Setten \emph{et~al.}(2007)Van~Setten, Popa, De~Wijs, and
  Brocks]{van2007electronic}
M.~Van~Setten, V.~Popa, G.~De~Wijs and G.~Brocks, \emph{Phys. Rev. B}, 2007,
  \textbf{75}, 035204\relax
\mciteBstWouldAddEndPuncttrue
\mciteSetBstMidEndSepPunct{\mcitedefaultmidpunct}
{\mcitedefaultendpunct}{\mcitedefaultseppunct}\relax
\EndOfBibitem
\bibitem[Li \emph{et~al.}(2011)Li, Peng, Zhou, and Wan]{li2011research}
C.~Li, P.~Peng, D.~Zhou and L.~Wan, \emph{Int. J. Hydrog. Energy}, 2011,
  \textbf{36}, 14512--14526\relax
\mciteBstWouldAddEndPuncttrue
\mciteSetBstMidEndSepPunct{\mcitedefaultmidpunct}
{\mcitedefaultendpunct}{\mcitedefaultseppunct}\relax
\EndOfBibitem
\bibitem[George and Saxena(2010)]{george2010structural}
L.~George and S.~K. Saxena, \emph{Int. J. Hydrog. Energy}, 2010, \textbf{35},
  5454--5470\relax
\mciteBstWouldAddEndPuncttrue
\mciteSetBstMidEndSepPunct{\mcitedefaultmidpunct}
{\mcitedefaultendpunct}{\mcitedefaultseppunct}\relax
\EndOfBibitem
\bibitem[y~De~Dompablo and Ceder(2004)]{y2004first}
M.~A. y~De~Dompablo and G.~Ceder, \emph{J Alloys Compd.}, 2004, \textbf{364},
  6--12\relax
\mciteBstWouldAddEndPuncttrue
\mciteSetBstMidEndSepPunct{\mcitedefaultmidpunct}
{\mcitedefaultendpunct}{\mcitedefaultseppunct}\relax
\EndOfBibitem
\bibitem[Peng and Chen(2008)]{peng2008ammonia}
B.~Peng and J.~Chen, \emph{Energy Environ. Sci}, 2008, \textbf{1},
  479--483\relax
\mciteBstWouldAddEndPuncttrue
\mciteSetBstMidEndSepPunct{\mcitedefaultmidpunct}
{\mcitedefaultendpunct}{\mcitedefaultseppunct}\relax
\EndOfBibitem
\bibitem[Bluhm \emph{et~al.}(2006)Bluhm, Bradley, Butterick, Kusari, and
  Sneddon]{bluhm2006amineborane}
M.~E. Bluhm, M.~G. Bradley, R.~Butterick, U.~Kusari and L.~G. Sneddon, \emph{J.
  Am. Chem. Soc}, 2006, \textbf{128}, 7748--7749\relax
\mciteBstWouldAddEndPuncttrue
\mciteSetBstMidEndSepPunct{\mcitedefaultmidpunct}
{\mcitedefaultendpunct}{\mcitedefaultseppunct}\relax
\EndOfBibitem
\bibitem[Miranda and Ceder(2007)]{miranda2007ab}
C.~R. Miranda and G.~Ceder, \emph{J. Chem. Phys.}, 2007, \textbf{126},
  184703\relax
\mciteBstWouldAddEndPuncttrue
\mciteSetBstMidEndSepPunct{\mcitedefaultmidpunct}
{\mcitedefaultendpunct}{\mcitedefaultseppunct}\relax
\EndOfBibitem
\bibitem[Li \emph{et~al.}(2014)Li, Yang, Chen, and Shore]{li2014ammonia}
H.~Li, Q.~Yang, X.~Chen and S.~G. Shore, \emph{J. Organomet. Chem.}, 2014,
  \textbf{751}, 60--66\relax
\mciteBstWouldAddEndPuncttrue
\mciteSetBstMidEndSepPunct{\mcitedefaultmidpunct}
{\mcitedefaultendpunct}{\mcitedefaultseppunct}\relax
\EndOfBibitem
\bibitem[Lin and Mao(2014)]{lin2014high}
Y.~Lin and W.~L. Mao, \emph{Chinese Sci. Bull.}, 2014, \textbf{59},
  5235--5240\relax
\mciteBstWouldAddEndPuncttrue
\mciteSetBstMidEndSepPunct{\mcitedefaultmidpunct}
{\mcitedefaultendpunct}{\mcitedefaultseppunct}\relax
\EndOfBibitem
\bibitem[Ravindran \emph{et~al.}(2002)Ravindran, Vajeeston, Vidya, Kjekshus,
  and Fjellv{\aa}g]{ravindran2002violation}
P.~Ravindran, P.~Vajeeston, R.~Vidya, A.~Kjekshus and H.~Fjellv{\aa}g,
  \emph{Phys. Rev. Lett.}, 2002, \textbf{89}, 106403\relax
\mciteBstWouldAddEndPuncttrue
\mciteSetBstMidEndSepPunct{\mcitedefaultmidpunct}
{\mcitedefaultendpunct}{\mcitedefaultseppunct}\relax
\EndOfBibitem
\bibitem[Vajeeston \emph{et~al.}(2004)Vajeeston, Ravindran, Fjellv{\aa}g, and
  Kjekshus]{vajeeston2004search}
P.~Vajeeston, P.~Ravindran, H.~Fjellv{\aa}g and A.~Kjekshus, \emph{Phys. Rev.
  B}, 2004, \textbf{70}, 014107\relax
\mciteBstWouldAddEndPuncttrue
\mciteSetBstMidEndSepPunct{\mcitedefaultmidpunct}
{\mcitedefaultendpunct}{\mcitedefaultseppunct}\relax
\EndOfBibitem
\bibitem[Siegel \emph{et~al.}(2007)Siegel, Wolverton, and
  Ozoli{\c{n}}{\v{s}}]{siegel2007thermodynamic}
D.~J. Siegel, C.~Wolverton and V.~Ozoli{\c{n}}{\v{s}}, \emph{Physical review
  B}, 2007, \textbf{76}, 134102\relax
\mciteBstWouldAddEndPuncttrue
\mciteSetBstMidEndSepPunct{\mcitedefaultmidpunct}
{\mcitedefaultendpunct}{\mcitedefaultseppunct}\relax
\EndOfBibitem
\bibitem[Z{\"u}ttel \emph{et~al.}(2003)Z{\"u}ttel, Rentsch, Fischer, Wenger,
  Sudan, Mauron, and Emmenegger]{zuttel2003hydrogen}
A.~Z{\"u}ttel, S.~Rentsch, P.~Fischer, P.~Wenger, P.~Sudan, P.~Mauron and
  C.~Emmenegger, \emph{Journal of Alloys and Compounds}, 2003, \textbf{356},
  515--520\relax
\mciteBstWouldAddEndPuncttrue
\mciteSetBstMidEndSepPunct{\mcitedefaultmidpunct}
{\mcitedefaultendpunct}{\mcitedefaultseppunct}\relax
\EndOfBibitem
\bibitem[Yang \emph{et~al.}(2010)Yang, Sudik, Wolverton, and
  Siegel]{yang2010high}
J.~Yang, A.~Sudik, C.~Wolverton and D.~J. Siegel, \emph{Chemical Society
  Reviews}, 2010, \textbf{39}, 656--675\relax
\mciteBstWouldAddEndPuncttrue
\mciteSetBstMidEndSepPunct{\mcitedefaultmidpunct}
{\mcitedefaultendpunct}{\mcitedefaultseppunct}\relax
\EndOfBibitem
\bibitem[Xiong \emph{et~al.}(2005)Xiong, Hu, Wu, Chen, Luo, Gross, and
  Wang]{xiong2005thermodynamic}
Z.~Xiong, J.~Hu, G.~Wu, P.~Chen, W.~Luo, K.~Gross and J.~Wang, \emph{Journal of
  Alloys and Compounds}, 2005, \textbf{398}, 235--239\relax
\mciteBstWouldAddEndPuncttrue
\mciteSetBstMidEndSepPunct{\mcitedefaultmidpunct}
{\mcitedefaultendpunct}{\mcitedefaultseppunct}\relax
\EndOfBibitem
\bibitem[Wolf \emph{et~al.}(2000)Wolf, Baumann, Baitalow, and
  Hoffmann]{wolf2000calorimetric}
G.~Wolf, J.~Baumann, F.~Baitalow and F.~Hoffmann, \emph{Thermochimica Acta},
  2000, \textbf{343}, 19--25\relax
\mciteBstWouldAddEndPuncttrue
\mciteSetBstMidEndSepPunct{\mcitedefaultmidpunct}
{\mcitedefaultendpunct}{\mcitedefaultseppunct}\relax
\EndOfBibitem
\bibitem[Hohenberg and Kohn(1964)]{hohenberg1964inhomogeneous}
P.~Hohenberg and W.~Kohn, \emph{Phys. Rev}, 1964, \textbf{136}, B864\relax
\mciteBstWouldAddEndPuncttrue
\mciteSetBstMidEndSepPunct{\mcitedefaultmidpunct}
{\mcitedefaultendpunct}{\mcitedefaultseppunct}\relax
\EndOfBibitem
\bibitem[Perdew \emph{et~al.}(1996)Perdew, Burke, and
  Ernzerhof]{perdew1996generalized}
J.~P. Perdew, K.~Burke and M.~Ernzerhof, \emph{Phys. Rev. Lett.}, 1996,
  \textbf{77}, 3865\relax
\mciteBstWouldAddEndPuncttrue
\mciteSetBstMidEndSepPunct{\mcitedefaultmidpunct}
{\mcitedefaultendpunct}{\mcitedefaultseppunct}\relax
\EndOfBibitem
\bibitem[Perdew(1986)]{perdew1986density}
J.~P. Perdew, \emph{Phys. Rev. B}, 1986, \textbf{33}, 8822\relax
\mciteBstWouldAddEndPuncttrue
\mciteSetBstMidEndSepPunct{\mcitedefaultmidpunct}
{\mcitedefaultendpunct}{\mcitedefaultseppunct}\relax
\EndOfBibitem
\bibitem[Kresse and Hafner(1993)]{kresse1993ab}
G.~Kresse and J.~Hafner, \emph{Phys. Rev. B}, 1993, \textbf{47}, 558\relax
\mciteBstWouldAddEndPuncttrue
\mciteSetBstMidEndSepPunct{\mcitedefaultmidpunct}
{\mcitedefaultendpunct}{\mcitedefaultseppunct}\relax
\EndOfBibitem
\bibitem[Dronskowski and Bloechl(1993)]{dronskowski1993crystal}
R.~Dronskowski and P.~E. Bloechl, \emph{J. Phys. Chem.}, 1993, \textbf{97},
  8617--8624\relax
\mciteBstWouldAddEndPuncttrue
\mciteSetBstMidEndSepPunct{\mcitedefaultmidpunct}
{\mcitedefaultendpunct}{\mcitedefaultseppunct}\relax
\EndOfBibitem
\bibitem[Dion \emph{et~al.}(2004)Dion, Rydberg, Schr{\"o}der, Langreth, and
  Lundqvist]{dion2004van}
M.~Dion, H.~Rydberg, E.~Schr{\"o}der, D.~C. Langreth and B.~I. Lundqvist,
  \emph{Phys. Rev. Lett.}, 2004, \textbf{92}, 246401\relax
\mciteBstWouldAddEndPuncttrue
\mciteSetBstMidEndSepPunct{\mcitedefaultmidpunct}
{\mcitedefaultendpunct}{\mcitedefaultseppunct}\relax
\EndOfBibitem
\bibitem[Rom{\'a}n-P{\'e}rez and Soler(2009)]{roman2009efficient}
G.~Rom{\'a}n-P{\'e}rez and J.~M. Soler, \emph{Phys. Rev. Lett.}, 2009,
  \textbf{103}, 096102\relax
\mciteBstWouldAddEndPuncttrue
\mciteSetBstMidEndSepPunct{\mcitedefaultmidpunct}
{\mcitedefaultendpunct}{\mcitedefaultseppunct}\relax
\EndOfBibitem
\bibitem[Klime{\v{s}} \emph{et~al.}(2009)Klime{\v{s}}, Bowler, and
  Michaelides]{klimevs2009chemical}
J.~Klime{\v{s}}, D.~R. Bowler and A.~Michaelides, \emph{J. Phys. Condens.
  Matter}, 2009, \textbf{22}, 022201\relax
\mciteBstWouldAddEndPuncttrue
\mciteSetBstMidEndSepPunct{\mcitedefaultmidpunct}
{\mcitedefaultendpunct}{\mcitedefaultseppunct}\relax
\EndOfBibitem
\bibitem[Lee \emph{et~al.}(2010)Lee, Murray, Kong, Lundqvist, and
  Langreth]{lee2010higher}
K.~Lee, {\'E}.~D. Murray, L.~Kong, B.~I. Lundqvist and D.~C. Langreth,
  \emph{Phys. Rev. B}, 2010, \textbf{82}, 081101\relax
\mciteBstWouldAddEndPuncttrue
\mciteSetBstMidEndSepPunct{\mcitedefaultmidpunct}
{\mcitedefaultendpunct}{\mcitedefaultseppunct}\relax
\EndOfBibitem
\bibitem[Klime{\v{s}} \emph{et~al.}(2011)Klime{\v{s}}, Bowler, and
  Michaelides]{klimevs2011van}
J.~Klime{\v{s}}, D.~R. Bowler and A.~Michaelides, \emph{Phys. Rev. B}, 2011,
  \textbf{83}, 195131\relax
\mciteBstWouldAddEndPuncttrue
\mciteSetBstMidEndSepPunct{\mcitedefaultmidpunct}
{\mcitedefaultendpunct}{\mcitedefaultseppunct}\relax
\EndOfBibitem
\bibitem[Thonhauser \emph{et~al.}(2007)Thonhauser, Cooper, Li, Puzder,
  Hyldgaard, and Langreth]{thonhauser2007van}
T.~Thonhauser, V.~R. Cooper, S.~Li, A.~Puzder, P.~Hyldgaard and D.~C. Langreth,
  \emph{Phys. Rev. B}, 2007, \textbf{76}, 125112\relax
\mciteBstWouldAddEndPuncttrue
\mciteSetBstMidEndSepPunct{\mcitedefaultmidpunct}
{\mcitedefaultendpunct}{\mcitedefaultseppunct}\relax
\EndOfBibitem
\bibitem[Grimme(2006)]{grimme2006semiempirical}
S.~Grimme, \emph{J. Comput. Chem.}, 2006, \textbf{27}, 1787--1799\relax
\mciteBstWouldAddEndPuncttrue
\mciteSetBstMidEndSepPunct{\mcitedefaultmidpunct}
{\mcitedefaultendpunct}{\mcitedefaultseppunct}\relax
\EndOfBibitem
\bibitem[Grimme \emph{et~al.}(2010)Grimme, Antony, Ehrlich, and
  Krieg]{grimme2010consistent}
S.~Grimme, J.~Antony, S.~Ehrlich and H.~Krieg, \emph{J. Chem. Phys.}, 2010,
  \textbf{132}, 154104\relax
\mciteBstWouldAddEndPuncttrue
\mciteSetBstMidEndSepPunct{\mcitedefaultmidpunct}
{\mcitedefaultendpunct}{\mcitedefaultseppunct}\relax
\EndOfBibitem
\bibitem[Grimme \emph{et~al.}(2011)Grimme, Ehrlich, and
  Goerigk]{grimme2011effect}
S.~Grimme, S.~Ehrlich and L.~Goerigk, \emph{J. Comput. Chem.}, 2011,
  \textbf{32}, 1456--1465\relax
\mciteBstWouldAddEndPuncttrue
\mciteSetBstMidEndSepPunct{\mcitedefaultmidpunct}
{\mcitedefaultendpunct}{\mcitedefaultseppunct}\relax
\EndOfBibitem
\bibitem[Tkatchenko and Scheffler(2009)]{tkatchenko2009accurate}
A.~Tkatchenko and M.~Scheffler, \emph{Phys. Rev. Lett.}, 2009, \textbf{102},
  073005\relax
\mciteBstWouldAddEndPuncttrue
\mciteSetBstMidEndSepPunct{\mcitedefaultmidpunct}
{\mcitedefaultendpunct}{\mcitedefaultseppunct}\relax
\EndOfBibitem
\bibitem[Tkatchenko \emph{et~al.}(2012)Tkatchenko, DiStasio~Jr, Car, and
  Scheffler]{tkatchenko2012accurate}
A.~Tkatchenko, R.~A. DiStasio~Jr, R.~Car and M.~Scheffler, \emph{Phys. Rev.
  Lett.}, 2012, \textbf{108}, 236402\relax
\mciteBstWouldAddEndPuncttrue
\mciteSetBstMidEndSepPunct{\mcitedefaultmidpunct}
{\mcitedefaultendpunct}{\mcitedefaultseppunct}\relax
\EndOfBibitem
\bibitem[Bucko \emph{et~al.}(2013)Bucko, Lebegue, Hafner, and
  Ángyán]{bucko2013improved}
T.~Bucko, S.~Lebegue, J.~Hafner and J.~G. Ángyán, \emph{J. Chem. Theory
  Comput.}, 2013, \textbf{9}, 4293--4299\relax
\mciteBstWouldAddEndPuncttrue
\mciteSetBstMidEndSepPunct{\mcitedefaultmidpunct}
{\mcitedefaultendpunct}{\mcitedefaultseppunct}\relax
\EndOfBibitem
\bibitem[Bu{\v{c}}ko \emph{et~al.}(2014)Bu{\v{c}}ko, Leb{\`e}gue,
  {\'A}ngy{\'a}n, and Hafner]{buvcko2014extending}
T.~Bu{\v{c}}ko, S.~Leb{\`e}gue, J.~G. {\'A}ngy{\'a}n and J.~Hafner, \emph{J.
  Chem. Phys.}, 2014, \textbf{141}, 034114\relax
\mciteBstWouldAddEndPuncttrue
\mciteSetBstMidEndSepPunct{\mcitedefaultmidpunct}
{\mcitedefaultendpunct}{\mcitedefaultseppunct}\relax
\EndOfBibitem
\bibitem[Bu{\v{c}}ko \emph{et~al.}(2016)Bu{\v{c}}ko, Leb{\`e}gue, Gould, and
  {\'A}ngy{\'a}n]{buvcko2016many}
T.~Bu{\v{c}}ko, S.~Leb{\`e}gue, T.~Gould and J.~G. {\'A}ngy{\'a}n, \emph{J.
  Phys. Condens. Matter}, 2016, \textbf{28}, 045201\relax
\mciteBstWouldAddEndPuncttrue
\mciteSetBstMidEndSepPunct{\mcitedefaultmidpunct}
{\mcitedefaultendpunct}{\mcitedefaultseppunct}\relax
\EndOfBibitem
\bibitem[Ambrosetti \emph{et~al.}(2014)Ambrosetti, Reilly, DiStasio~Jr, and
  Tkatchenko]{ambrosetti2014long}
A.~Ambrosetti, A.~M. Reilly, R.~A. DiStasio~Jr and A.~Tkatchenko, \emph{J.
  chem. phys.}, 2014, \textbf{140}, 18A508\relax
\mciteBstWouldAddEndPuncttrue
\mciteSetBstMidEndSepPunct{\mcitedefaultmidpunct}
{\mcitedefaultendpunct}{\mcitedefaultseppunct}\relax
\EndOfBibitem
\bibitem[Steinmann and Corminboeuf(2011)]{steinmann2011comprehensive}
S.~N. Steinmann and C.~Corminboeuf, \emph{J. Chem. Theory Comput.}, 2011,
  \textbf{7}, 3567--3577\relax
\mciteBstWouldAddEndPuncttrue
\mciteSetBstMidEndSepPunct{\mcitedefaultmidpunct}
{\mcitedefaultendpunct}{\mcitedefaultseppunct}\relax
\EndOfBibitem
\bibitem[Steinmann and Corminboeuf(2011)]{steinmann2011generalized}
S.~N. Steinmann and C.~Corminboeuf, \emph{J. chem. phys.}, 2011, \textbf{134},
  044117\relax
\mciteBstWouldAddEndPuncttrue
\mciteSetBstMidEndSepPunct{\mcitedefaultmidpunct}
{\mcitedefaultendpunct}{\mcitedefaultseppunct}\relax
\EndOfBibitem
\bibitem[Becke and Johnson(2005)]{becke2005exchange}
A.~D. Becke and E.~R. Johnson, \emph{J. chem. phys.}, 2005, \textbf{122},
  154104\relax
\mciteBstWouldAddEndPuncttrue
\mciteSetBstMidEndSepPunct{\mcitedefaultmidpunct}
{\mcitedefaultendpunct}{\mcitedefaultseppunct}\relax
\EndOfBibitem
\bibitem[Bultinck \emph{et~al.}(2007)Bultinck, Van~Alsenoy, Ayers, and
  Carb{\'o}-Dorca]{bultinck2007critical}
P.~Bultinck, C.~Van~Alsenoy, P.~W. Ayers and R.~Carb{\'o}-Dorca, \emph{J. chem.
  phys.}, 2007, \textbf{126}, 144111\relax
\mciteBstWouldAddEndPuncttrue
\mciteSetBstMidEndSepPunct{\mcitedefaultmidpunct}
{\mcitedefaultendpunct}{\mcitedefaultseppunct}\relax
\EndOfBibitem
\bibitem[Tiritiris and Schleid(2003)]{tiritiris2003dodekahydro}
I.~Tiritiris and T.~Schleid, \emph{Z. Anorg. Allg. Chem}, 2003, \textbf{629},
  1390--1402\relax
\mciteBstWouldAddEndPuncttrue
\mciteSetBstMidEndSepPunct{\mcitedefaultmidpunct}
{\mcitedefaultendpunct}{\mcitedefaultseppunct}\relax
\EndOfBibitem
\bibitem[B{\"o}ddeker \emph{et~al.}(1966)B{\"o}ddeker, Shore, and
  Bunting]{boddeker1966boron}
K.~W. B{\"o}ddeker, S.~G. Shore and R.~K. Bunting, \emph{J. Am. Chem. Soc.},
  1966, \textbf{88}, 4396--4401\relax
\mciteBstWouldAddEndPuncttrue
\mciteSetBstMidEndSepPunct{\mcitedefaultmidpunct}
{\mcitedefaultendpunct}{\mcitedefaultseppunct}\relax
\EndOfBibitem
\bibitem[Leavers \emph{et~al.}(1969)Leavers, Long, Shore, and
  Taylor]{leavers1969heat}
D.~Leavers, J.~Long, S.~G. Shore and W.~Taylor, \emph{Journal of the Chemical
  Society A: Inorganic, Physical, Theoretical}, 1969,  1580--1581\relax
\mciteBstWouldAddEndPuncttrue
\mciteSetBstMidEndSepPunct{\mcitedefaultmidpunct}
{\mcitedefaultendpunct}{\mcitedefaultseppunct}\relax
\EndOfBibitem
\bibitem[Sun \emph{et~al.}(2011)Sun, Wolverton, Akbarzadeh, and
  Ozolins]{sun2011first}
W.~Q. Sun, C.~Wolverton, A.~Akbarzadeh and V.~Ozolins, \emph{Phys. Rev. B},
  2011, \textbf{83}, 064112\relax
\mciteBstWouldAddEndPuncttrue
\mciteSetBstMidEndSepPunct{\mcitedefaultmidpunct}
{\mcitedefaultendpunct}{\mcitedefaultseppunct}\relax
\EndOfBibitem
\bibitem[Vinet \emph{et~al.}(1986)Vinet, Ferrante, Smith, and
  Rose]{vinet1986universal}
P.~Vinet, J.~Ferrante, J.~Smith and J.~Rose, \emph{J. Phys. C}, 1986,
  \textbf{19}, L467\relax
\mciteBstWouldAddEndPuncttrue
\mciteSetBstMidEndSepPunct{\mcitedefaultmidpunct}
{\mcitedefaultendpunct}{\mcitedefaultseppunct}\relax
\EndOfBibitem
\bibitem[Vinet \emph{et~al.}(1989)Vinet, Rose, Ferrante, and
  Smith]{vinet1989universal}
P.~Vinet, J.~H. Rose, J.~Ferrante and J.~R. Smith, \emph{J. Phys. Condens.
  Matter}, 1989, \textbf{1}, 1941\relax
\mciteBstWouldAddEndPuncttrue
\mciteSetBstMidEndSepPunct{\mcitedefaultmidpunct}
{\mcitedefaultendpunct}{\mcitedefaultseppunct}\relax
\EndOfBibitem
\bibitem[Birch(1947)]{birch1947finite}
F.~Birch, \emph{Phys. Rev.}, 1947, \textbf{71}, 809\relax
\mciteBstWouldAddEndPuncttrue
\mciteSetBstMidEndSepPunct{\mcitedefaultmidpunct}
{\mcitedefaultendpunct}{\mcitedefaultseppunct}\relax
\EndOfBibitem
\bibitem[Murnaghan(1944)]{murnaghan1944compressibility}
F.~Murnaghan, \emph{Proc Natl Acad Sci}, 1944, \textbf{30}, 244--247\relax
\mciteBstWouldAddEndPuncttrue
\mciteSetBstMidEndSepPunct{\mcitedefaultmidpunct}
{\mcitedefaultendpunct}{\mcitedefaultseppunct}\relax
\EndOfBibitem
\bibitem[Ravindran \emph{et~al.}(2006)Ravindran, Vajeeston, Vidya,
  Fjellv{\aa}g, and Kjekshus]{ravindran2006modeling}
P.~Ravindran, P.~Vajeeston, R.~Vidya, H.~Fjellv{\aa}g and A.~Kjekshus, \emph{J.
  Power Sources}, 2006, \textbf{159}, 88--99\relax
\mciteBstWouldAddEndPuncttrue
\mciteSetBstMidEndSepPunct{\mcitedefaultmidpunct}
{\mcitedefaultendpunct}{\mcitedefaultseppunct}\relax
\EndOfBibitem
\bibitem[Vajeeston \emph{et~al.}(2004)Vajeeston, Ravindran, Vidya,
  Fjellv{\aa}g, and Kjekshus]{vajeeston2004design}
P.~Vajeeston, P.~Ravindran, R.~Vidya, H.~Fjellv{\aa}g and A.~Kjekshus,
  \emph{Cryst. Growth Des.}, 2004, \textbf{4}, 471--477\relax
\mciteBstWouldAddEndPuncttrue
\mciteSetBstMidEndSepPunct{\mcitedefaultmidpunct}
{\mcitedefaultendpunct}{\mcitedefaultseppunct}\relax
\EndOfBibitem
\bibitem[King-Smith and Vanderbilt(1993)]{king1993theory}
R.~King-Smith and D.~Vanderbilt, \emph{Physical Review B}, 1993, \textbf{47},
  1651\relax
\mciteBstWouldAddEndPuncttrue
\mciteSetBstMidEndSepPunct{\mcitedefaultmidpunct}
{\mcitedefaultendpunct}{\mcitedefaultseppunct}\relax
\EndOfBibitem
\bibitem[Bader(1990)]{bader1990atoms}
R.~F. Bader, \emph{Atoms in molecules: a quantum theory, vol 22, International
  series of monographs on chemistry}, 1990\relax
\mciteBstWouldAddEndPuncttrue
\mciteSetBstMidEndSepPunct{\mcitedefaultmidpunct}
{\mcitedefaultendpunct}{\mcitedefaultseppunct}\relax
\EndOfBibitem
\bibitem[Henkelman \emph{et~al.}(2006)Henkelman, Arnaldsson, and
  J{\'o}nsson]{henkelman2006fast}
G.~Henkelman, A.~Arnaldsson and H.~J{\'o}nsson, \emph{Computational Materials
  Science}, 2006, \textbf{36}, 354--360\relax
\mciteBstWouldAddEndPuncttrue
\mciteSetBstMidEndSepPunct{\mcitedefaultmidpunct}
{\mcitedefaultendpunct}{\mcitedefaultseppunct}\relax
\EndOfBibitem
\bibitem[Ramzan \emph{et~al.}(2009)Ramzan, Silvearv, Blomqvist, Scheicher,
  Lebegue, and Ahuja]{ramzan2009structural}
M.~Ramzan, F.~Silvearv, A.~Blomqvist, R.~H. Scheicher, S.~Lebegue and R.~Ahuja,
  \emph{Phys. Rev. B}, 2009, \textbf{79}, 132102\relax
\mciteBstWouldAddEndPuncttrue
\mciteSetBstMidEndSepPunct{\mcitedefaultmidpunct}
{\mcitedefaultendpunct}{\mcitedefaultseppunct}\relax
\EndOfBibitem
\bibitem[Hamilton \emph{et~al.}(2009)Hamilton, Baker, Staubitz, and
  Manners]{hamilton2009b}
C.~W. Hamilton, R.~T. Baker, A.~Staubitz and I.~Manners, \emph{Chem. Soc.
  Rev.}, 2009, \textbf{38}, 279--293\relax
\mciteBstWouldAddEndPuncttrue
\mciteSetBstMidEndSepPunct{\mcitedefaultmidpunct}
{\mcitedefaultendpunct}{\mcitedefaultseppunct}\relax
\EndOfBibitem
\bibitem[Matus \emph{et~al.}(2009)Matus, Grant, Nguyen, and
  Dixon]{matus2009fundamental}
M.~H. Matus, D.~J. Grant, M.~T. Nguyen and D.~A. Dixon, \emph{J. Phys. Chem.
  C}, 2009, \textbf{113}, 16553--16560\relax
\mciteBstWouldAddEndPuncttrue
\mciteSetBstMidEndSepPunct{\mcitedefaultmidpunct}
{\mcitedefaultendpunct}{\mcitedefaultseppunct}\relax
\EndOfBibitem
\bibitem[Mulliken(1955)]{mulliken1955rs}
R.~Mulliken, \emph{J. Chem. Phys.}, 1955, \textbf{23}, 1833\relax
\mciteBstWouldAddEndPuncttrue
\mciteSetBstMidEndSepPunct{\mcitedefaultmidpunct}
{\mcitedefaultendpunct}{\mcitedefaultseppunct}\relax
\EndOfBibitem
\end{mcitethebibliography}
\bibliographystyle{rsc} 

\end{document}